\documentclass[superscriptaddress,amsmath,amssymb,nofootinbib,eqsecnum,a4paper,secnumarabic,preprintnumbers,11pt]{revtex4}         
\usepackage[pdftex]{graphicx}
\usepackage{hyperref}     
\usepackage{slashed}
\usepackage{color}
\usepackage{braket}
\usepackage{cancel}

\newcommand{\tr}{\mbox{tr}}

\usepackage{titlesec}     
\newcommand*{\justifyheading}{\raggedright}
\titleformat{\chapter}[display]
  {\normalfont\huge\bfseries\justifyheading}{\chaptertitlename\ \thechapter}
  {20pt}{\Huge}
\titleformat{\section}
  {\normalfont\Large\bfseries\justifyheading}{\thesection}{1em}{}
\titleformat{\subsection}
  {\normalfont\large\bfseries\justifyheading}{\thesubsection}{1em}{}
\titleformat{\subsubsection}
  {\normalfont\bfseries\justifyheading}{\thesubsection}{1em}{}
\usepackage[english]{babel}

\preprint{IPMU17-0184}
\preprint{TU-1055}
\begin{document}

\title{Model independent evaluation of the Wilson coefficient \\ of the Weinberg operator in QCD}
\author{Tomohiro Abe}
\affiliation{
  Institute for Advanced Research, Nagoya University,
  Furo-cho Chikusa-ku, Nagoya, Aichi, 464-8602 Japan
}
\affiliation{
  Kobayashi-Maskawa Institute for the Origin of Particles and the
  Universe, Nagoya University,
  Furo-cho Chikusa-ku, Nagoya, Aichi, 464-8602 Japan
}

\author{Junji Hisano}
\affiliation{
  Kobayashi-Maskawa Institute for the Origin of Particles and the
  Universe, Nagoya University,
  Furo-cho Chikusa-ku, Nagoya, Aichi, 464-8602 Japan
}
\affiliation{
Department of Physics, Nagoya University,
Furo-cho Chikusa-ku, Nagoya, Aichi, 464-8602 Japan
}
\affiliation{
  Kavli IPMU (WPI), UTIAS, University of Tokyo, Kashiwa, Chiba 277-8584, Japan
}

\author{Ryo Nagai}
\affiliation{Department of Physics, Tohoku University, Sendai, Miyagi 980-8578, Japan}

\begin{abstract}
We derive a Wilson coefficient of a CP-violating purely gluonic dimension-6 operator 
called the Weinberg operator ($GG\tilde{G}$) 
generated by a scalar and two fermions at the two-loop level. 
We do not specify the representation of SU(3)$_c$ for the scalar and the fermions,
and thus our result can be applied to a variety of models beyond the standard model. 
We estimate the nucleon EDMs induced by the Weinberg operator in some examples and discuss the importance of measuring EDMs. It is found that future measurements of the EDMs can probe physics at higher energy scale beyond the reach of collider experiments.
\end{abstract}

\maketitle

\section{Introduction}
Measurements of electric dipole moments (EDMs) are very powerful for exploring physics beyond the standard model (SM).
The SM predicts small values of EDMs.
Its prediction for the neutron EDM is 
$d_n \simeq 10^{-32}$ e cm~\cite{PhysLett.B109.215, PhysLett.109B.490, PhysLett.B197.556, NUPHA.B244.77, hep-ph/0504231},
and for the electron EDM is $d_e \leq 10^{-38}$ e cm~\cite{Pospelov:1991zt, hep-ph/0504231}. %
It is much smaller than the current upper bounds,
$|d_n| < 2.9 \times 10^{-26}$ e cm (90\% CL)~\cite{hep-ex/0602020},
and $|d_e| < 8.7 \times 10^{-29}$ e cm (90\% CL)~\cite{1310.7534}.
On the other hand, models beyond the SM often have new CP violation sources,
and they can predict larger values of EDMs compared to the SM.
Therefore observation of EDMs is equivalent to a discovery of physics beyond the SM.
Moreover, EDMs can probe physics at higher energy scale beyond the reach of collider experiments.
The current data of the Large Hadron Collider (LHC) experiment imply that 
the scale of new physics is higher than ${\cal O}(1)$~TeV. 
Thus the importance of measuring EDMs is increasing.
This situation motivates us to evaluate EDMs in a variety of models with small theoretical uncertainties.

Weinberg pointed out that 
a pure gluonic dimension-6 operator, $GG\tilde{G}$, is a source of CP violation in QCD, 
and it affects to the EDM of neutrons~\cite{PRLTA.63.2333}.
We call the operator as the Weinberg operator.
In renormalizable models, two-loop diagrams are at the leading order for the Weinberg operator.
Full two-loop calculations have done for some models:
two-Higgs doublet models with CP violation~\cite{PHRVA.D41.999},
the quark-squark-gluino diagram in MSSM~\cite{PHLTA.B237.216},
and
Left-Right symmetric models~\cite{PHRVA.D42.867, PHLTA.B249.467}.
Besides full two-loop calculations, it has been shown that the Weinberg operator is induced 
as a result of integrating out heavy colored particles which carry chromo-EDMs (cEDMs)~\cite{PRLTA.64.1709,
PHLTA.B241.584,PHLTA.B241.589,PHLTA.B242.239}. 
The calculation is based on the effective theory, and the calculation is simpler than the full two-loop calculation.
However, it is justified only if the following two conditions are satisfied;
(i) One of the particles in the loop is lighter than the others and
(ii) the lightest particle is color non-singlet. 
It is useful to derive more general formulae to evaluate contributions to the Weinberg operator
from new physics in general setup.

In this paper, we calculate two-loop diagrams that contain one scalar and two fermions
without specifying their color representations,
and derive a general formula for the Wilson coefficient of the Weinberg operator in the system.
Our result can be applied to a variety models.
We also compare our full two-loop calculation to the result derived in the effective theory.
It is shown that the calculation based on the effective theory can overestimate the contribution
if the mass difference among the particles in the loop is small.
For the application of our general formula, we estimate the nucleon EDMs in some examples. 
Comparing our results with the latest upper bound on nucleon EDMs, 
we survey the parameter region which is consistent with current experimental results. 
We also discuss the impact of measuring EDMs by showing the parameter region which can be covered by the future observations of the EDMs.

The rest of this paper is organized as follows.
At first, we introduce our setup and fix the notation for calculating the Wilson coefficient of the Weinberg operator in section \ref{sec:setup}. In section \ref{sec:diagrams}, we perform full two-loop calculation and derive a general formula for the Wilson coefficient of the Weinberg operator. 
The result of the two-loop calculation is summarized in section \ref{sec:result}. In section \ref{sec:effective}, we discuss the relationship between our result and the result based on the effective theoretical approach. We estimate the nucleon EDMs in some examples and discuss the importance of measuring EDMs in section \ref{sec:Numeric}. Finally, we summarize our discussion in section \ref{sec:summary}.

\section{Setup}
\label{sec:setup}
\begin{figure}[tb]
\includegraphics[width=0.40\hsize]{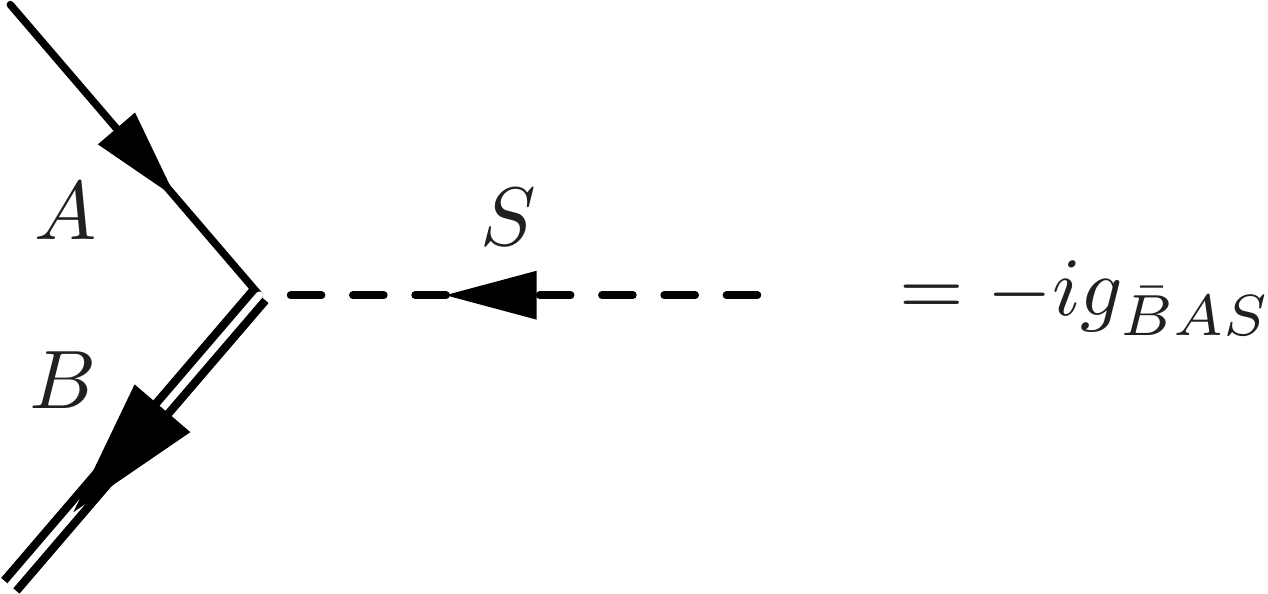}
\qquad \qquad \qquad
\includegraphics[width=0.40\hsize]{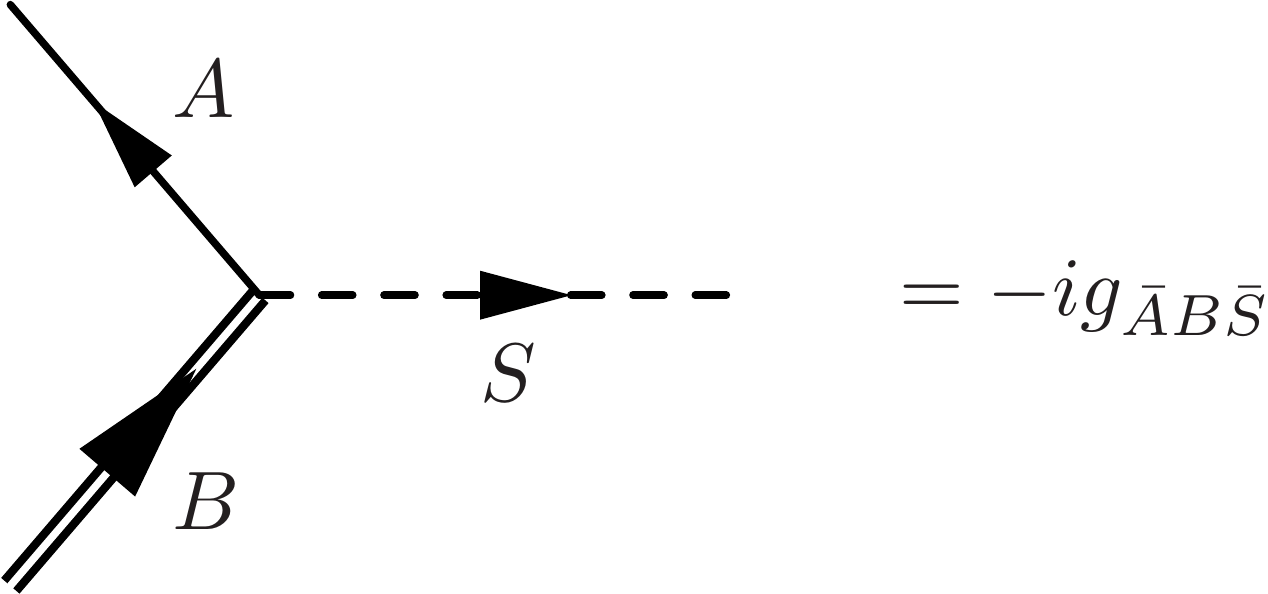}
\caption{
Feynman rules.
}
\label{fig:FMrules}
\end{figure}

We consider two fermions ($A$ and $B$) and a scalar field ($S$) 
which have the following interactions,
\begin{align}
 {\cal L} \supset& 
- \bar{\psi}_B g_{\bar{B}AS} \psi_A S
- \bar{\psi}_A g_{\bar{A}B\bar{S}} \psi_B S^*,
\end{align}
where
\begin{align}
 g_{\bar{B}AS} =& X_{\bar{B}AS} (s + \gamma_5 a), \\
 g_{\bar{A}B\bar{S}} =& X^\dagger_{\bar{A}B\bar{S}} (s^* - \gamma_5 a^*).
\end{align}
All the indices of SU(3)$_c$ representations are described by $X$.
Explicit expressions for some examples are shown in Table.~\ref{tab:X}.
$s$ and $a$ are complex numbers.
The Feynman rules are shown in Fig.~\ref{fig:FMrules}.
\begin{table}[tbp]
\centering
\caption{Group factors in some representations.
}
\label{tab:X}
\begin{tabular}{|c|c|c|c|c|c|} 
\hline
$(A,B,S)$ & $\psi_A \ \psi_B \ S$ & $X_{\bar{B}AS}$ & $XT_AT_AX^\dagger$ &  $XT_A X^\dagger T_B$ & $X X^\dagger T_B T_B$\\ \hline\hline
$(3,3,1)$ & $(\psi_A)^a \ (\psi_B)^b \ S$ & $\delta^b_a$ & $\frac{1}{2}$ & $\frac{1}{2}$  & $\frac{1}{2}$  \\ \hline
$(3,1,\bar{3})$ & $(\psi_A)^a \ (\psi_B) \ S_i$ & $\delta^i_a$ & $\frac{1}{2}$ & $0$  & $0$  \\ \hline
$(1,3,3)$ & $(\psi_A) \ (\psi_B)^b \ S^i$ & $\delta^b_i$ & 0 & $0$  & $\frac{1}{2}$  \\ \hline
\hline
$(\bar{6},1,6)$ & $(\psi_A)_{ab} \ (\psi_B) \ S^{ij}$ & $\frac{\delta^a_i \delta^b_j + \delta^a_j \delta^b_i}{2}$ & $\frac{5}{2}$ & $0$ & $0$  \\ \hline
$(6,6,1)$ & $(\psi_A)^{ij} \ (\psi_B)^{kl} \ S$ & $\frac{\delta^k_i \delta^l_j + \delta^k_j \delta^l_i}{2}$ & $\frac{5}{2}$ & $\frac{5}{2}$ & $\frac{5}{2}$ \\ \hline
\hline
$(3,\bar{3},3)$ & $(\psi_A)^i \ (\psi_B)_j \ S^k$ & $\epsilon_{ijk}$ & $1$ & $\frac{1}{2}$  & $1$  \\ \hline
$(\bar{3},3,6)$ & $(\psi_A)_a \ (\psi_B)^b \ S^{ij}$ & $\frac{\delta^b_i \delta^a_j + \delta^b_j \delta^a_i}{2}$ & $1$ & $-\frac{1}{4}$  & $1$  \\ \hline
$(3,3,8)$ & $(\psi_A)^i \ (\psi_B)^j \ (S^a T^a)$ & $(T^a)^j_{\ i}$ & $\frac{2}{3}$ & $-\frac{1}{12}$  & $\frac{2}{3}$  \\ \hline
$(3,8,\bar{3})$ & $(\psi_A)^i \ (\psi_B^a T^a) \ S_j $ & $(T^a)^j_{\ i}$ & $\frac{2}{3}$ & $\frac{3}{4}$  & $\frac{3}{2}$  \\ \hline
\hline
\end{tabular}
\end{table}

In this setup, we will calculated the Wilson coefficient of the Weinberg operator, $C_G$, defined
\begin{align}
 {\cal L} \supset g_s^3 C_G {\cal O},
\end{align}
where
\begin{align}
{\cal O} 
=&
 -\frac{1}{3} f^{abc} G^a_{\mu \nu} G^b{}^{\nu}_{\ \rho} \tilde{G}^{c \rho \mu} 
,
\label{eq:weinberg_operator}
\end{align}
and $g_s$ is the SU(3)$_c$ gauge coupling defined in the covariant derivative by
\begin{align}
 D_\mu = \partial_\mu + i g_s T^a A^a_\mu.
\end{align}
In Eq.~\eqref{eq:weinberg_operator}, 
$f^{abc}$ is the structure constant of SU(3)$_c$, 
and $\tilde{G}^{c \rho \mu} = \frac{1}{2} \epsilon^{\rho \mu \alpha \beta} G^c_{\alpha \beta}$ 
with $\epsilon^{0123} = +1$.

\section{Evaluation of diagrams}
\label{sec:diagrams}
The leading order contributions to the Weinberg operator are given by two-loop diagrams.
We use the Fock-Schwinger gauge for the external gluon fields.
The technical details of the Fock-Schwinger gauge is reviewed, for example, in Refs.~\cite{Novikov:1983gd, 1007.2601}.

In addition to the Weinberg operator ($GG\tilde{G}$), 
we also need to calculate operators which consist of $G_{\mu \nu}$, $\tilde{G}_{\mu \nu}$,
and two covariant derivatives, such as $DGD\tilde{G}$.
This is because 
the external gluon fields are described by covariant derivatives and gluon field strength in
the Fock-Schwinger gauge,
and commutation of two covariant derivatives is equivalent to gluon field strength.

\subsection{$GG\tilde{G}$ from $DGD\tilde{G}$ terms}
\label{sec:GGtildeG_and_DGDtildeG}
\begin{figure}[tb]
\includegraphics[width=0.20\hsize]{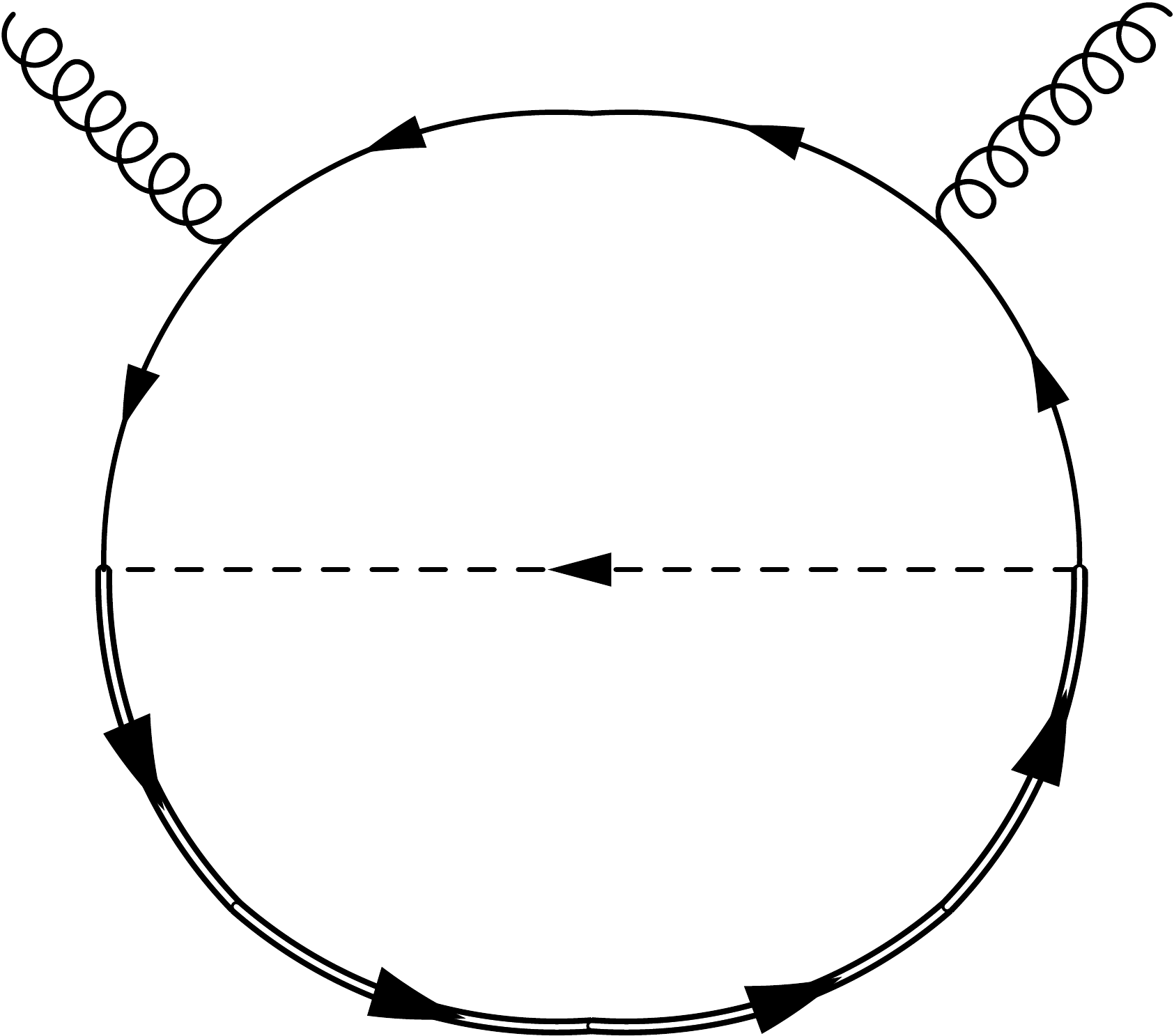}
\qquad \qquad
\includegraphics[width=0.15\hsize]{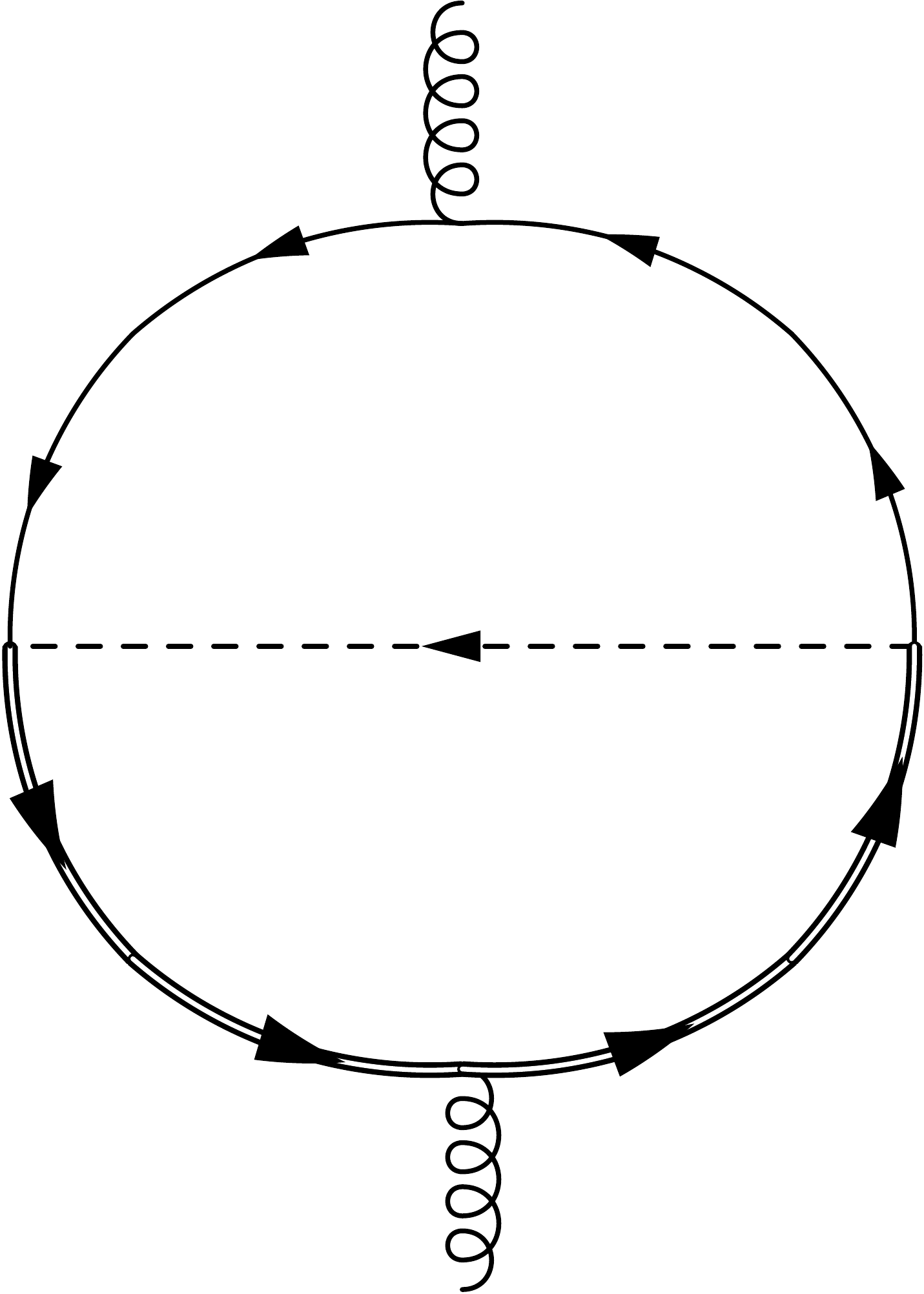}
\qquad \qquad
\includegraphics[width=0.20\hsize]{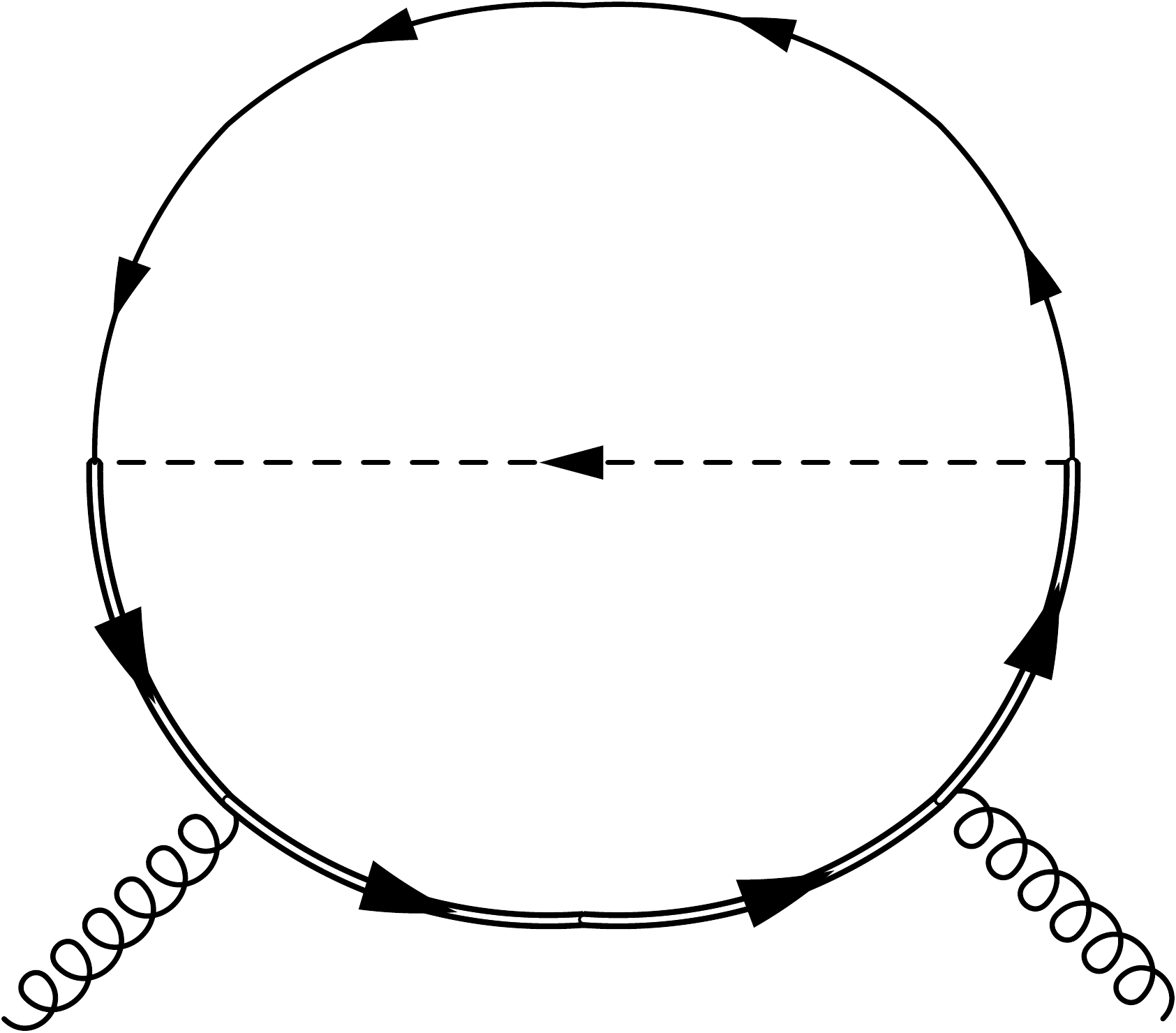}
\caption{
The diagrams that generate $DGD\tilde{G}$.
We call each diagram from the left to the right as (2,0), (1,1), and (0,2), respectively.
}
\label{fig:Diagrams}
\end{figure}

We calculate $GG\tilde{G}$ from $DGD\tilde{G}$ terms. We evaluate diagrams with two external gluon fields.
Diagrams should have at least one gluon field from fermion line, otherwise terms with $\gamma_5$ all vanish.
In addition, diagrams with one gluon field from the scalar field also vanish.
See appendix \ref{app:GluonsFromScalar} for the detail.
Figure~\ref{fig:Diagrams} shows the diagrams we have to evaluate.
We call each diagram from left to right as (2,0), (1,1), and (0,2), respectively.

In the calculation, we obtain terms that are combination of $X$ and SU(3)$_c$ generator, such as
$\left(X_{\bar{B}AS} (T^a)_{AA'} (T^b)_{A'A''} X^\dagger_{\bar{A}''B\bar{S}}\right)$.
Here we denote the SU(3)$_c$ indices of $T^a$ and $T^b$ symbolically. 
Since $X$ consists of $\epsilon$ tensors and Kronecker deltas, 
$\left(X_{\bar{B}AS} (T^a)_{AA'} (T^b)_{A'A''} X^\dagger_{\bar{A}''B\bar{S}}\right)$ is proportional to $\tr(T^a T^b) \propto \delta^{ab}$.
Other similar terms are also proportional to $\delta^{ab}$.
Thus we can introduce the following notations.
\begin{align}
 \left(X_{\bar{B}AS} (T^a)_{AA'} (T^b)_{A'A''} X^\dagger_{\bar{A}''B\bar{S}}\right)
\equiv&
 \left(X T_A T_A X^\dagger \right) \delta^{ab}
,\\
 \left(X_{\bar{B}AS} (T^a)_{AA'} X^\dagger_{\bar{A}'B'\bar{S}} (T^b)_{B'B} \right)
\equiv&
 \left(X T_A X^\dagger T_B \right) \delta^{ab}
,\\
 \left(X_{\bar{B}AS} X^\dagger_{\bar{A}B'\bar{S}} (T^a)_{B'B''} (T^b)_{B''B} \right)
\equiv&
 \left(X X^\dagger T_B T_B \right) \delta^{ab}
.
\end{align}
where the subscript $F=A,B$ denotes that $T_{F}$ is the generator for the fermion field $F$.
The values of $\left(X T_A T_A X^\dagger \right)$ etc. depend on the representations of $A$, $B$, and $S$.
We show some examples in Table~\ref{tab:X}.
We use \texttt{FeynCalc}~\cite{Mertig:1990an,1601.01167} to evaluate traces with $\gamma$-matrices.
We also use the following identities that can be proved by using Jacobi identities and Bianchi identities.
\begin{align}
D_{\alpha_1} G^a_{\alpha \mu} D_{\beta_1} G^a_{\beta \nu}
\left(
 g^{\alpha \beta} \epsilon^{\alpha_1 \beta_1 \mu \nu}
+g^{\alpha \beta_1} \epsilon^{\alpha_1 \beta \mu \nu}
+g^{\alpha_1 \beta} \epsilon^{\alpha \beta_1 \mu \nu}
+g^{\alpha_1 \beta_1} \epsilon^{\alpha_1 \beta_1 \mu \nu}
\right)
=&
 27 g_s {\cal O}
,\\
G^a_{\alpha \mu} D_{\beta_2}  D_{\beta_1} G^a_{\beta \nu}
\left(
 g^{\beta \beta_1} \epsilon^{\alpha \beta_2 \mu \nu}
+g^{\beta \beta_2} \epsilon^{\alpha \beta_1 \mu \nu}
+g^{\beta_1 \beta_2} \epsilon^{\alpha \beta \mu \nu}
\right)
=&
- 18 g_s {\cal O}
,\\
(D_{\alpha_2} D_{\alpha_1} G^a_{\alpha \mu}) G^a_{\beta \nu}
\left(
 g^{\alpha \alpha_1} \epsilon^{\alpha_2 \beta \mu \nu}
+g^{\alpha \alpha_2} \epsilon^{\alpha_1 \beta \mu \nu}
+g^{\alpha_1 \alpha_2} \epsilon^{\alpha \beta \mu \nu}
\right)
=&
- 18 g_s {\cal O}
.
\end{align}
Since the subdiagrams of the (2,0) and (0,2) contain UV-divergences,
we need counterterms to renormalize them during our calculation.
However, as we will see in later, 
the final result after summing all of diagrams is independent from the UV-divergence.

We find the following $DGD\tilde{G}$ terms.
\begin{align}
 \text{(2,0)-diagram}
=&
i \frac{6}{(4\pi)^4} g_s^3 \left(X T_A T_A X^\dagger \right) {\cal O} 
\nonumber\\
&
\times
\Biggl[
i \frac{\text{Re}(sa^*)}{2} 
h_1(m_A^2, m_B^2, m_S^2)
+ \text{Im}(sa^*) 
h_2(m_A^2, m_B^2, m_S^2)
\Biggr]
,
\\
 \text{(0,2)-diagram}
=&
i \frac{6}{(4\pi)^4} g_s^3 \left(X X^\dagger T_B T_B \right) {\cal O} 
\nonumber\\
&
\times
\Biggl[
i \frac{\text{Re}(sa^*)}{2} 
h_1(m_B^2, m_A^2, m_S^2)
+ \text{Im}(sa^*) 
h_2(m_B^2, m_A^2, m_S^2)
\Biggr]
,
\\
 \text{(1,1)-diagram}
=&
i \frac{6}{(4\pi)^4} g_s^3 \left(X T_A X^\dagger T_B \right) {\cal O}
\nonumber\\
&
\times
\Biggl[
i \frac{\text{Re}(sa^*)}{2} 
h_3(m_A^2, m_B^2, m_S^2)
+ \text{Im}(sa^*)
h_4(m_A^2, m_B^2, m_S^2)
+ (A \leftrightarrow B)
\Biggr]
,
\label{eq:1-1_diagram}
\end{align}
where
\begin{align}
h_1(m_A^2, m_B^2, m_S^2)
=&
\left( \frac{i}{(4\pi)^2} \right)^{-1} 
\int_\ell \frac{\ell^2}{(\ell^2 - m_A^2)^4}
(\tilde{B}_0 + \tilde{B}_1)_{(\ell^2, m_S^2, m_B^2)}
\label{eq:h1}
,\\
h_2(m_A^2, m_B^2, m_S^2)
=&
\left( \frac{i}{(4\pi)^2} \right)^{-1} 
\int_\ell \frac{m_A m_B}{(\ell^2 - m_A^2)^4} 
\tilde{B}_0{}_{(\ell^2, m_S^2, m_B^2)} 
,\\
h_3(m_A^2, m_B^2, m_S^2)
=&
\left( \frac{i}{(4\pi)^2} \right)^{-1} 
\int_\ell 
\frac{\ell^2}{(\ell^2 - m_B^2)^3} 
\frac{\partial (B_0 + B_1)_{(\ell^2, m_S^2, m_A^2)}}{\partial m_A^2}
,\\
h_4(m_A^2, m_B^2, m_S^2)
=&
\left( \frac{i}{(4\pi)^2} \right)^{-1} 
\int_\ell 
\frac{m_A m_B}{(\ell^2 - m_B^2)^3} \frac{\partial B_0{}_{(\ell^2, m_S^2, m_A^2)}}{\partial m_A^2}
\label{eq:h4}
,\\
 B_0{}_{(\ell^2, m_S^2, m_A^2)}
=&
\left( \frac{i}{(4\pi)^2} \right)^{-1} 
\int_q \frac{1}{[q^2 - m_S^2][(q+\ell)^2 - m_A^2]}
 ,\\
\ell^{\mu} B_1{}_{(\ell^2, m_S^2, m_A^2)}
 =&
\left( \frac{i}{(4\pi)^2} \right)^{-1} 
  \int_q \frac{q^{\mu}}{[q^2 - m_S^2][(q+\ell)^2 - m_A^2]}
,
\end{align}
where $\int_{\ell} = \int \frac{d^4 \ell}{(2\pi)^4}$,
and 
$\tilde{B}_i = B_i + $(counter terms).
For example, 
$\tilde{B}_i{}_{(\ell^2, m_S^2, m_B^2)} = B_i{}_{(\ell^2, m_S^2, m_B^2)} - B_i{}_{(m_A^2, m_S^2, m_B^2)}$
with the on-shell renormalization condition.
However, the contributions from counterterms vanish after we add other diagrams discussed in the next section.
In the calculation for Eq.~\eqref{eq:1-1_diagram}, we have used relations given in Eqs.~\eqref{eq:A1} and \eqref{eq:A2}.

\subsection{Diagrams with three external gluon fields}
\begin{figure}[tb]
\includegraphics[width=0.20\hsize]{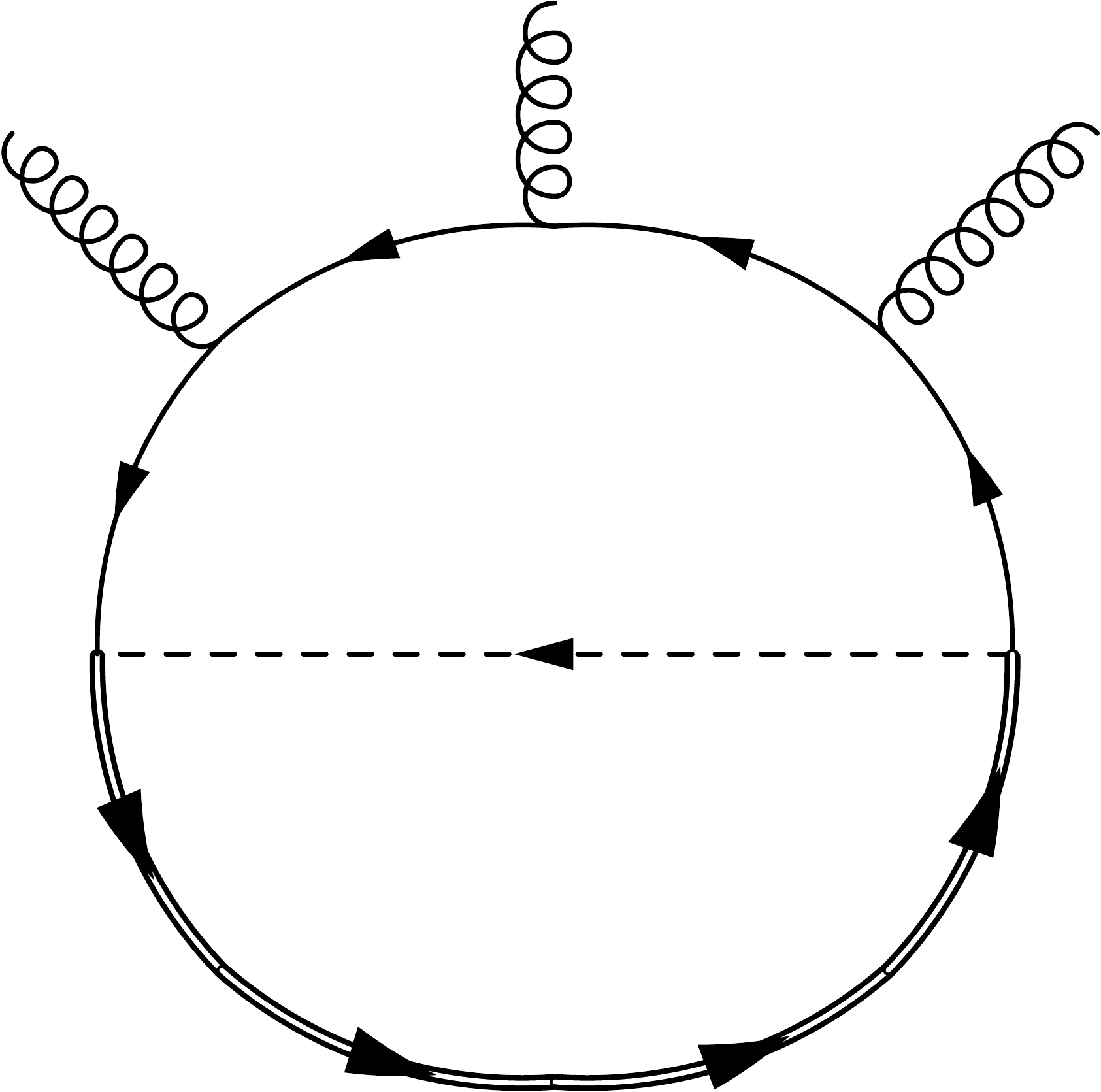}
\quad \quad
\includegraphics[width=0.20\hsize]{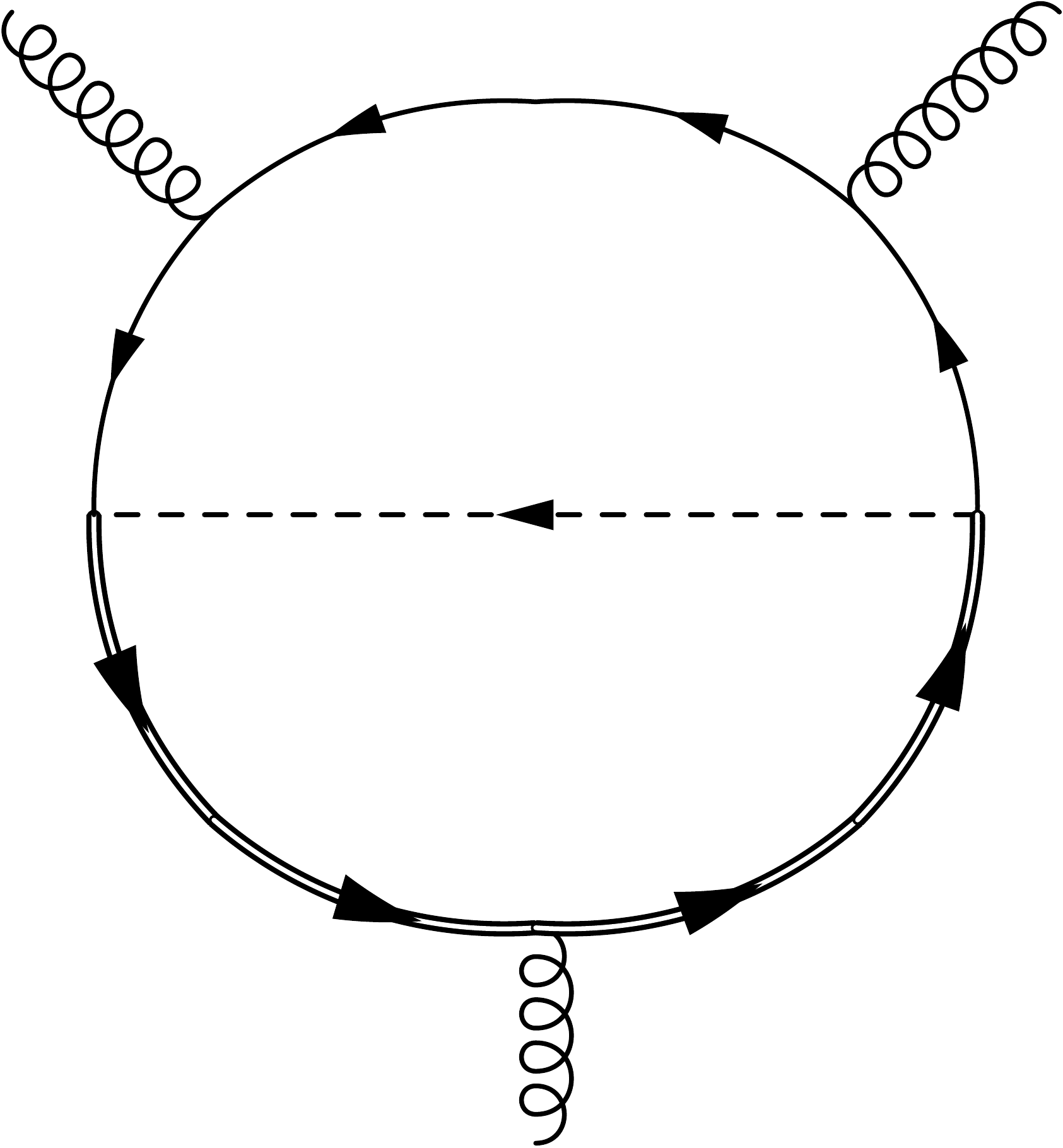}
\quad \quad
\includegraphics[width=0.20\hsize]{./21.pdf}
\quad \quad
\includegraphics[width=0.20\hsize]{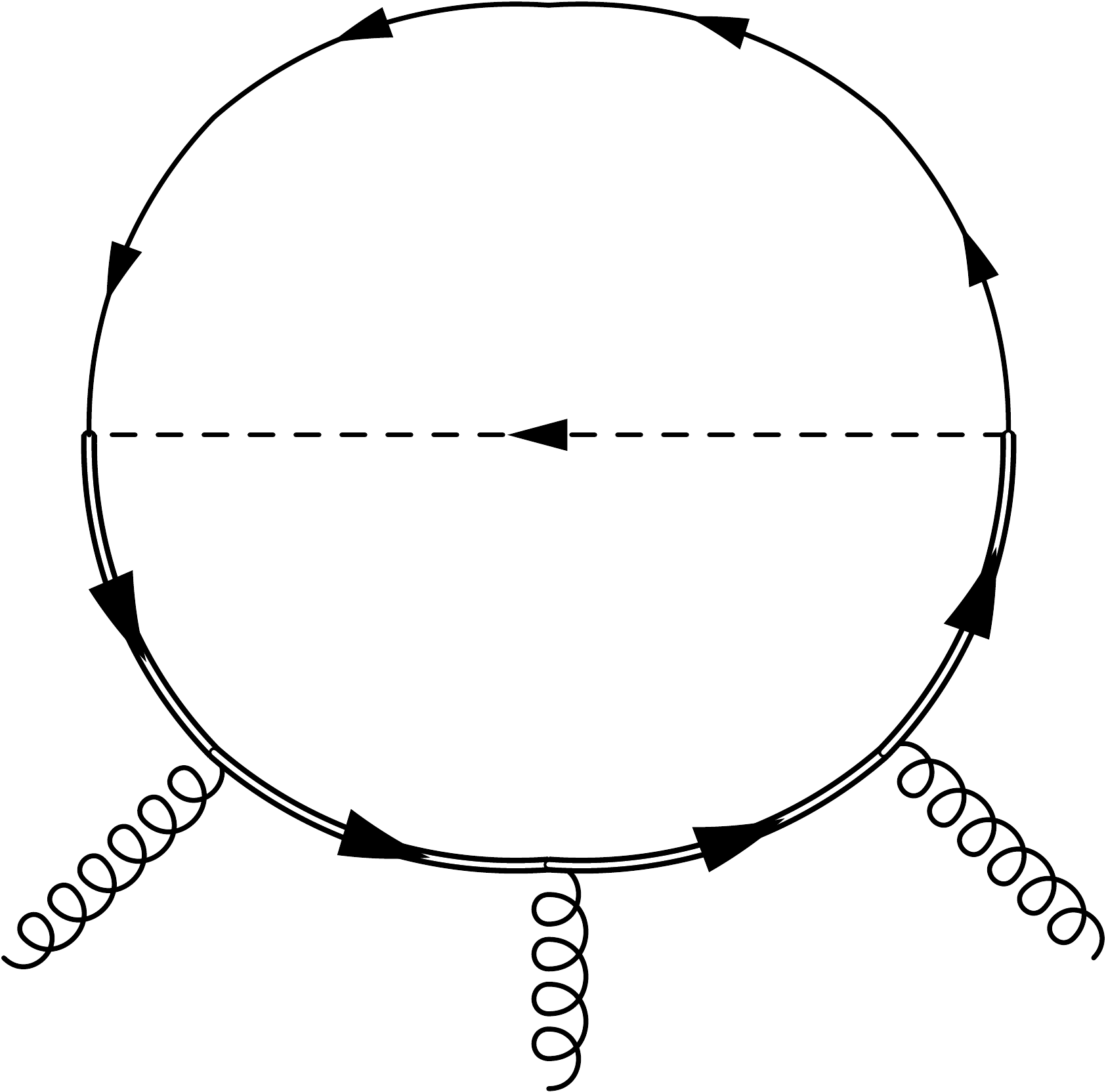}
\caption{
The diagrams that generate $GG\tilde{G}$.
We call each diagram from the left to the right as (3,0), (2,1), (1,2), and (0,3), respectively.
}
\label{fig:Diagrams_3g}
\end{figure}

We evaluate diagrams with three external gluon fields to obtain $GG\tilde{G}$ terms.
Because of the same reason as in $DGD\tilde{G}$ terms, 
diagrams should have at least two gluon field from fermion line. 
In addition, diagrams with one gluon field from the scalar field vanish
thanks to a virtue of the Fock-Schwinger gauge.
See appendix \ref{app:GluonsFromScalar} for the detail.
The diagrams we have to evaluate are shown in Fig.~\ref{fig:Diagrams_3g}.
We call each diagram from left to right as (3,0), (2,1), (1,2), and (0,3), respectively.

There are some terms that consist of $X$ and SU(3)$_c$ generators.
Direct calculations shows that they are proportional to the structure constant of SU(3)$_c$.
We can introduce the following notations.
\begin{align}
X_{\bar{B}AS} (T^a)_{AA'} (T^b)_{A'A''} (T^c)_{A''A'''} X^\dagger_{\bar{A'''}B\bar{S}}
\equiv& 
(X T_A T_A T_A X^\dagger) i f^{abc}
,\\
X_{\bar{B}AS} (T^a)_{AA'} (T^b)_{A'A''} X^\dagger_{\bar{A''}B'\bar{S}} (T^c)_{B'B}
\equiv& 
(X T_A T_A X^\dagger T_B) i f^{abc}
,\\
X_{\bar{B}AS} (T^a)_{AA'} X^\dagger_{\bar{A'}B'\bar{S}} (T^b)_{B'B''} (T^c)_{B''B}
\equiv& 
(X T_A X^\dagger T_B T_B) i f^{abc}
,\\
X_{\bar{B}AS} X^\dagger_{\bar{A}B'\bar{S}} (T^b)_{B'B''} (T^b)_{B''B'''} (T^c)_{B'''B}
\equiv& 
(X X^\dagger T_B T_B T_B) i f^{abc}
.
\end{align}
We find relations
\begin{align}
 \left( X T_A T_A T_A X^\dagger \right) {\cal O}
=&
 \frac{1}{2} \left( X T_A T_A X^\dagger \right) {\cal O}
,\\
 \left( X T_A T_A X^\dagger T_B \right) {\cal O}
=&
 \frac{1}{2} \left( X T_A X^\dagger T_B \right) {\cal O}
,\\
 \left( X T_A X^\dagger T_B T_B \right) {\cal O}
=&
 \frac{1}{2} \left( X T_A X^\dagger T_B \right) {\cal O}
,\\
 \left( X X^\dagger T_B T_B T_B \right) {\cal O}
=&
 \frac{1}{2} \left( X X^\dagger T_B T_B \right) {\cal O}
.
\end{align}
It is easy to prove these relations. For example,
\begin{align}
  \left( X T_A T_A T_A X^\dagger \right) (-6{\cal O})
=&
 -i \left( X T^a T^b T^c X^\dagger \right) G^a_{\mu \nu} G^b{}^\nu_{\ \rho} G^c_{\alpha \beta} \epsilon^{\rho \mu \alpha \beta}
\nonumber\\
=&
-\frac{i}{2} \left( X [T^a, T^b] T^c X^\dagger \right) G^a_{\mu \nu} G^b{}^\nu_{\ \rho} G^c_{\alpha \beta} \epsilon^{\rho \mu \alpha \beta}
\nonumber\\
=&
\frac{1}{2} f^{abd} \left( X T^d T^c X^\dagger \right) G^a_{\mu \nu} G^b{}^\nu_{\ \rho} G^c_{\alpha \beta} \epsilon^{\rho \mu \alpha \beta}
\nonumber\\
=&
\frac{1}{2} \left( X T_A T_A X^\dagger \right) (-6 {\cal O})
.
\label{eq:TTTtoTT}
\end{align}
Here we used a fact that 
$G^a_{\mu \nu} G^b{}^\nu_{\ \rho} \epsilon^{\rho \mu \alpha \beta}$ is anti-symmetric under exchange of $a$ and $b$.
Using the relations, we find 
\begin{align}
 \text{(3,0)-diagram}
=&
-i \frac{6}{(4\pi)^4} g_s^3 \left( X T_A T_A X^\dagger \right) {\cal O}
\nonumber\\
&
\quad
\times
\Biggl[
i\frac{\text{Re}(sa^*)}{2} 
h_1(m_A^2, m_B^2, m_S^2)
\nonumber\\
&
\qquad \quad
+ \text{Im}(sa^*)
\left(
h_2(m_A^2, m_B^2, m_S^2)
+
m_A m_B
f_1(m_A^2, m_B^2, m_S^2)
\right)
\Biggr]
,
\\
 \text{(0,3)-diagram}
=&
-i \frac{6}{(4\pi)^4} g_s^3 \left( X X^\dagger T_B T_B \right) {\cal O}
\nonumber\\
&
\quad
\times
\Biggl[
i \frac{\text{Re}(sa^*)}{2}
h_1(m_B^2, m_A^2, m_S^2)
\nonumber\\
&
\qquad \quad
+ \text{Im}(sa^*)
\left(
h_2(m_B^2, m_A^2, m_S^2)
+
m_A m_B
f_1(m_B^2, m_A^2, m_S^2)
\right)
\Biggr]
,
\\
\text{(2,1)-diagram}
+
 \text{(1,2)-diagram}
=&
-i \frac{6}{(4\pi)^4} g_s^3 \left( X T_A X^\dagger T_B \right) {\cal O}
\nonumber\\
&
\quad
\times
\Biggl[
 i\frac{\text{Re}(sa^*)}{2}
h_3(m_A^2, m_B^2, m_S^2)
\nonumber\\
&
\qquad \quad
+ \text{Im}(sa^*)
\left(
h_4(m_A^2, m_B^2, m_S^2)
+
m_A m_B
f_2(m_A^2, m_B^2, m_S^2)
\right)
\nonumber\\
&
\qquad \quad
+
(A \leftrightarrow B)
\Biggr]
,
\end{align}
where $h_i$ are defined in Eqs.~\eqref{eq:h1}--\eqref{eq:h4}, and 
\begin{align}
f_1(m_A^2, m_B^2, m_S^2)
\equiv&
\left( \frac{i}{(4\pi)^2} \right)^{-1}
\int_\ell 
 \frac{\ell^2}{(\ell^2 - m_A^2)^4} 
\frac{\partial B_0{}_{(\ell^2, m_S^2, m_B^2)}}{\partial \ell^2}
\nonumber\\
=&
\int_0^\infty d\ell_E^2
\int_0^1 dz
\frac{-\ell_E^4 z (1-z)}{\left( m_S^2 z + m_B^2 (1-z) + \ell_E^2 z (1-z) \right) \left( \ell_E^2 + m_A^2 \right)^4}
\label{eq:f1}
,\\
f_2(m_A^2, m_B^2, m_S^2)
\equiv&
\left( \frac{i}{(4\pi)^2} \right)^{-1}
\int_\ell 
 \frac{\ell^2}{(\ell^2 - m_A^2)^4} 
\frac{\partial B_0{}_{(\ell^2, m_S^2, m_B^2)}}{\partial m_B^2}
\nonumber\\
=&
\int_0^\infty d\ell_E^2
\int_0^1 dz
\frac{\ell_E^4 (1-z)}{\left( m_S^2 z + m_B^2 (1-z) + \ell_E^2 z (1-z) \right) \left( \ell_E^2 + m_A^2 \right)^4}
\label{eq:f2}
.
\end{align}
Similar to (2,0) and (0,2)-diagrams, (3,0) and (0,3)-diagrams contain UV-divergences,
and thus $h_1$ and $h_2$ contain counterterms.
In the derivation of $f_1$, we have used a relation given in Eq.~\eqref{eq:id_for_f1}.

\section{Results}
\label{sec:result}
We sum up all of the diagrams calculated in Sec.~\ref{sec:diagrams},
and find the following coefficient of the Weinberg operator.
\begin{align}
C_G
=&
-\frac{1}{(4\pi)^4} 6 
\text{Im}(sa^*) m_A m_B 
\nonumber\\
&
\times
\Biggl\{
\left(X T_A T_A X^\dagger \right) 
f_1(m_A^2, m_B^2, m_S^2)
+
\left(X X^\dagger T_B T_B \right) 
f_1(m_B^2, m_A^2, m_S^2)
\nonumber\\
&
\qquad 
+
\left(X T_A X^\dagger T_B \right)
\biggl[
f_2(m_A^2, m_B^2, m_S^2) + f_2(m_B^2, m_A^2, m_S^2)
\biggr]
\Biggr\}
,
\label{eq:our_result}
\end{align}
where $f_1$ and $f_2$ are defined in Eqs.~\eqref{eq:f1} and \eqref{eq:f2}.
From the definitions, we see that $f_1 < 0$ and $f_2 >0$.

If we use the same interaction term twice, 
we have to multiply 1/2 by Eq.~\eqref{eq:our_result}.
For instance, if $\psi_A=\psi_B$ and $S^* = S$, then Eq.~\eqref{eq:our_result}
is modified with extra 1/2 as follows.
\begin{align}
C_G
=&
-\frac{1}{2} \frac{1}{(4\pi)^4} 6 
\text{Im}(sa^*) m^2_A  
\nonumber\\
&
\times
\Biggl\{
\left(X T_A T_A X^\dagger \right) 
f_1(m_A^2, m_A^2, m_S^2)
+
\left(X X^\dagger T_A T_A \right) 
f_1(m_A^2, m_A^2, m_S^2)
\nonumber\\
&
\qquad 
+2
\left(X T_A X^\dagger T_A \right)
\biggl[
f_2(m_A^2, m_A^2, m_S^2)
\biggr]
\Biggr\}
.
\label{eq:our_result_1/2}
\end{align}

We make two comments. First one is that the cancellation of $\text{Re}(sa^*)$ terms. 
This cancellation is expected because these terms break hermicity of the Lagrangian. 
The second comment is the cancellation of the contributions from the counterterms,
which are in $h_1$ and $h_2$. Thus the result is independent from the renormalization
conditions for the subdiagrams in (2,0), (0,2), (3,0), and (0,3)-diagrams.

For numerical evaluations, we need to calculate group factors such as $X T_A T_A X^\dagger$.
Table~\ref{tab:X} shows some examples.
Note that 
if there is an SU(3)$_c$ singlet field, then
the non-zero factors are $N(r)$ that is defined through $\tr(T^a T^b) = N(r) \delta^{ab}$.


Our result is model independent in the sense that it can be applied to any models that generate the Weinberg operator from the diagrams shown in Fig.~\ref{fig:Diagrams}. 
For example, we can apply Eq.~\eqref{eq:our_result} to models that predict CP-violating quark-squark-gluino coupling by taking $A =q$, $B = \tilde{g}$, and $S = \tilde{q}$. 
Another example is a CP-violating $tth$-coupling, where we can use Eq.~\eqref{eq:our_result_1/2} by taking $A = B = t$ and $S = h$. 
We have checked that our result is consistent with the results reported in Refs.~\cite{PRLTA.63.2333,Ellis:2008zy,Jung:2013hka}.

\section{A view point from effective theory}
\label{sec:effective}
In this section, we discuss our result from a view point of effective theory.
For $m_A \ll m_B, m_S$,
the Weinberg operator can be 
generated from the chromo-electric dipole moment (cEDM) of the particle $A$ 
that is generated by the heavy particles ($B$ and $S$).
The authors in Refs.~\cite{Chang:1991ry, Chang:1992vs}
have calculated the Weinberg operator in the framework of effective theory assuming the existence of the cEDM 
at the UV cutoff scale for a fermion in the SU(3)$_c$ fundamental representation, and they have found
\begin{align}
 C_g^{eff.} =& \frac{d^c}{32\pi^2 m_A},
\label{eq:CGeff}
\end{align}
where $d^c$ is defined through
\begin{align}
{\cal L} \supset& \bar{Q} \left( -\frac{i}{2} g_s d^c \gamma_5 \sigma^{\mu \nu} G_{\mu \nu} \right) Q. 
\end{align}
Here $Q$ is in the SU(3)$_c$ fundamental representation.
In the following, we show that our full two-loop result agree with $C_g^{eff.}$ in $m_A \ll m_B, m_S$ regime.

\begin{figure}[tb]
\includegraphics[width=0.40\hsize]{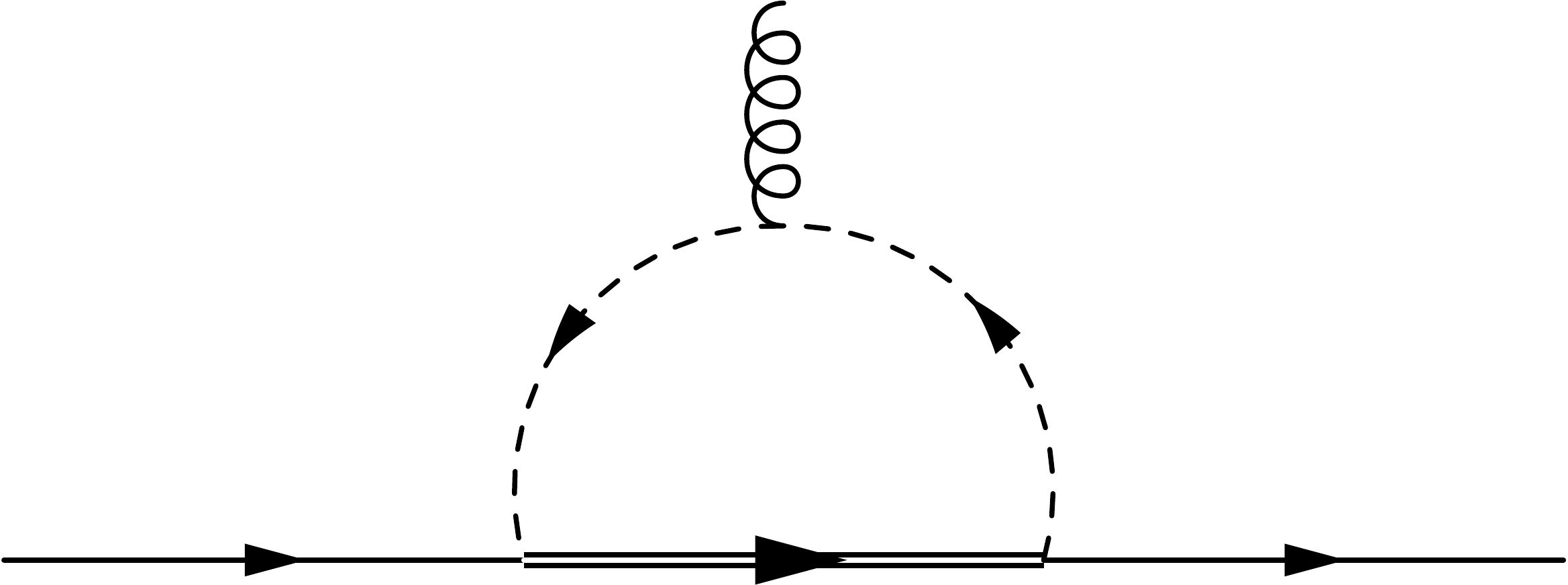}
\qquad \qquad \qquad
\includegraphics[width=0.40\hsize]{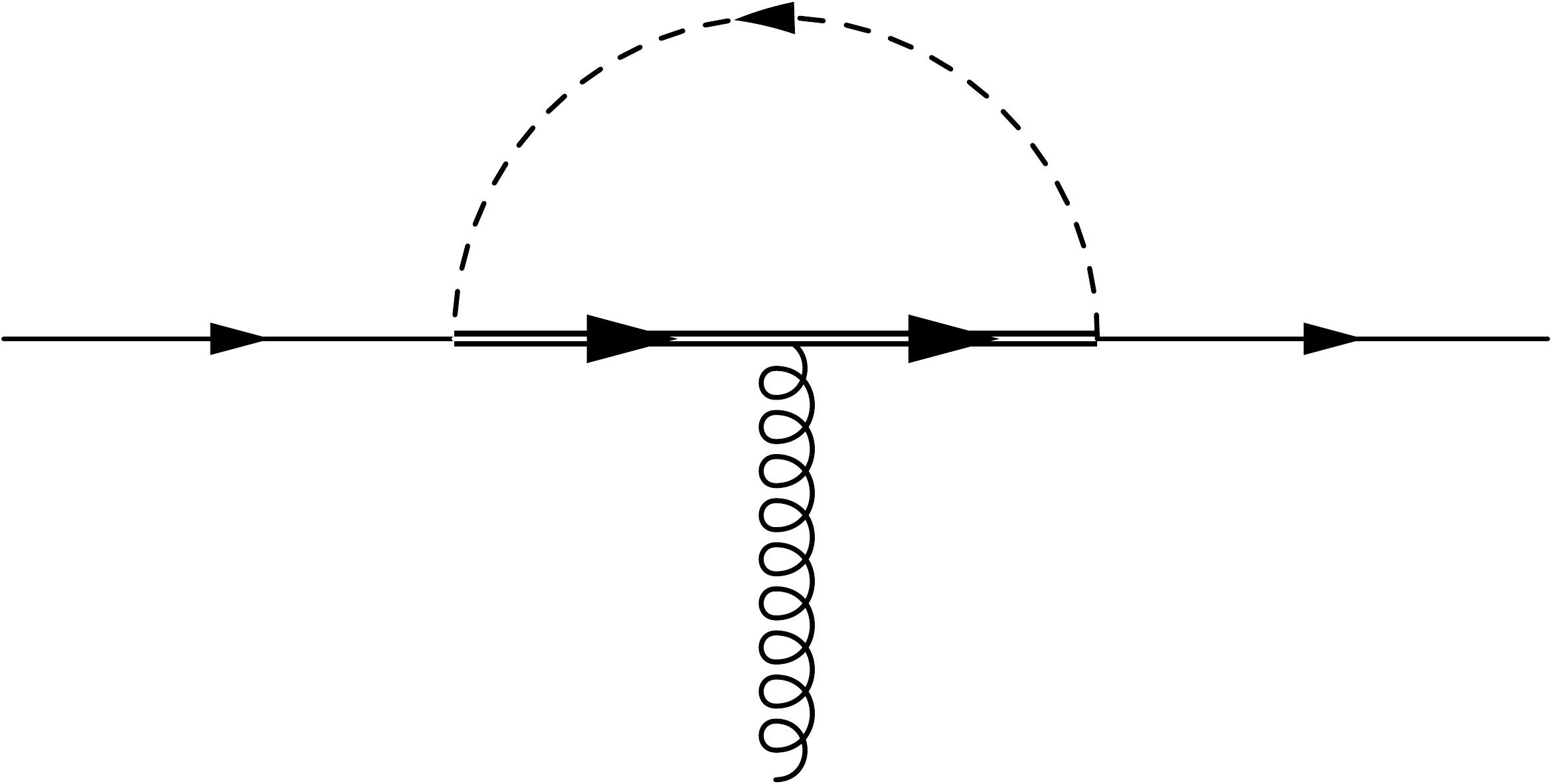}
\caption{The leading order diagrams for the cEDM of $A$.}
\label{fig:cEDM}
\end{figure}

In order to compare our result given in Eq.~\eqref{eq:our_result}
with Eq.~\eqref{eq:CGeff},
we need to calculate the cEDM of $A$, which is generated by the diagrams shown in Fig.~\ref{fig:cEDM}.\footnote{
We calculate the diagrams shown in Fig.~\ref{fig:cEDM} with the 't Hooft Feynman gauge, not with the Fock-Schwinger gauge.}
We introduce the following notations,
\begin{align}
\left( X^\dagger_{\bar{A}BS} (T^a)_{S'S} X_{\bar{B}A'S'} \right)
\equiv&
\left( X^\dagger T_S X\right) (T^a)_{AA'}
,\\
\left( X^\dagger_{\bar{A}BS} (T^a)_{BB'} X_{\bar{B'}A'S} \right)
\equiv&
\left( X^\dagger T_B X\right) (T^a)_{AA'}
\label{eq:XBX}
.
\end{align}
We denote the contribution to the cEDM from the diagram in the left (right) panel of Fig.~\ref{fig:cEDM} as
$d^c_S$ ($d^c_B$).
We find
\begin{align}
 d^c_S =&
- \frac{1}{(4\pi)^2} 2\text{Im}(s a^*) m_B
\left( X^\dagger T_S X\right)  
\int_0^1 dz \frac{z(1-z)}{m_S^2 z + m_B^2 (1-z) - m_A^2 z(1-z)}
\label{eq:dc_S}
,\\ 
 d^c_B =&
- \frac{1}{(4\pi)^2} 2\text{Im}(s a^*) m_B
\left( X^\dagger T_B X\right)  
\int_0^1 dz \frac{(1-z)^2}{m_S^2 z + m_B^2 (1-z) - m_A^2 z(1-z)}
\label{eq:dc_B}
.
\end{align}
The result depends on $T_S$ as we can see from Eq.~\eqref{eq:dc_S}.
However, $T_A$, $T_B$, and $T_S$ must be related to each other by the gauge invariance,
and thus $T_S$ can be written by $T_A$, $T_B$, and $X$. 
We find 
\begin{align}
 X^\dagger T_S X =&
X^\dagger T_B X - \frac{1}{N(r_A)} (X T_A T_A X^\dagger)
\label{eq:TS_vs_TA_vs_TB}
,\\
 X^\dagger T_B X =&
 \frac{1}{N(r_A)} (X T_A X^\dagger T_B)
\label{eq:TS_vs_TA_vs_TB_2}
.
\end{align}
We take $N(r_A) = 1/2$
because Eq.~\eqref{eq:CGeff} is derived with a fermion in the SU(3)$_c$ fundamental representation
in Refs.~\cite{Chang:1991ry, Chang:1992vs}.
A derivation of Eq.~\eqref{eq:TS_vs_TA_vs_TB} is given in Appendix~\ref{sec:TS_vs_TA_vs_TB}.
Eq.~\eqref{eq:TS_vs_TA_vs_TB_2} is derived by multiplying $(T^b)_{A'A}$ and Eq.~\eqref{eq:XBX}.
Using these relations, we can eliminate $T_S$ from Eq.~\eqref{eq:dc_S}, 
and results can be expressed only by $X T_A T_A X^\dagger$ and $X T_A X^\dagger T_B$.
We find
\begin{align}
 C_g^{eff.} =& 
-
\frac{1}{(4\pi)^4} 6 \text{Im}(s a^*) m_A m_B
\nonumber\\
&
\times
\Biggl(
- (X T_A T_A X^\dagger)
\frac{1}{3 m_A^2}
\int_0^1 dz \frac{z(1-z)}{m_S^2 z + m_B^2 (1-z) - m_A^2 z(1-z)}
\nonumber\\
& \qquad
+
 (X T_A X^\dagger T_B)
\frac{1}{3 m_A^2}
\int_0^1 dz \frac{(1-z)}{m_S^2 z + m_B^2 (1-z) - m_A^2 z(1-z)}
\Biggr)
.
\label{eq:ourCGeff}
\end{align}

\begin{figure}[tb]
\includegraphics[width=.49\hsize]{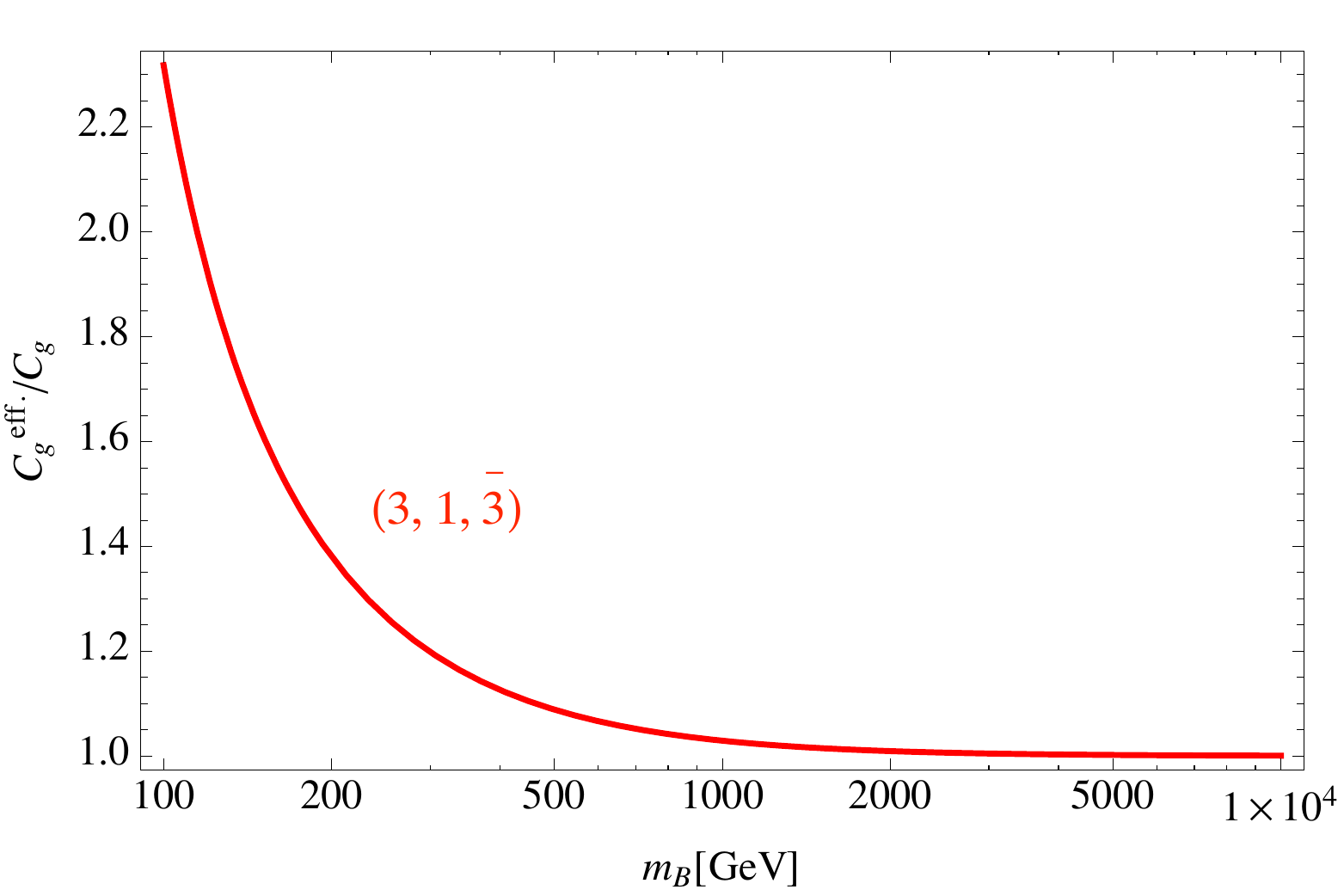}
\includegraphics[width=.49\hsize]{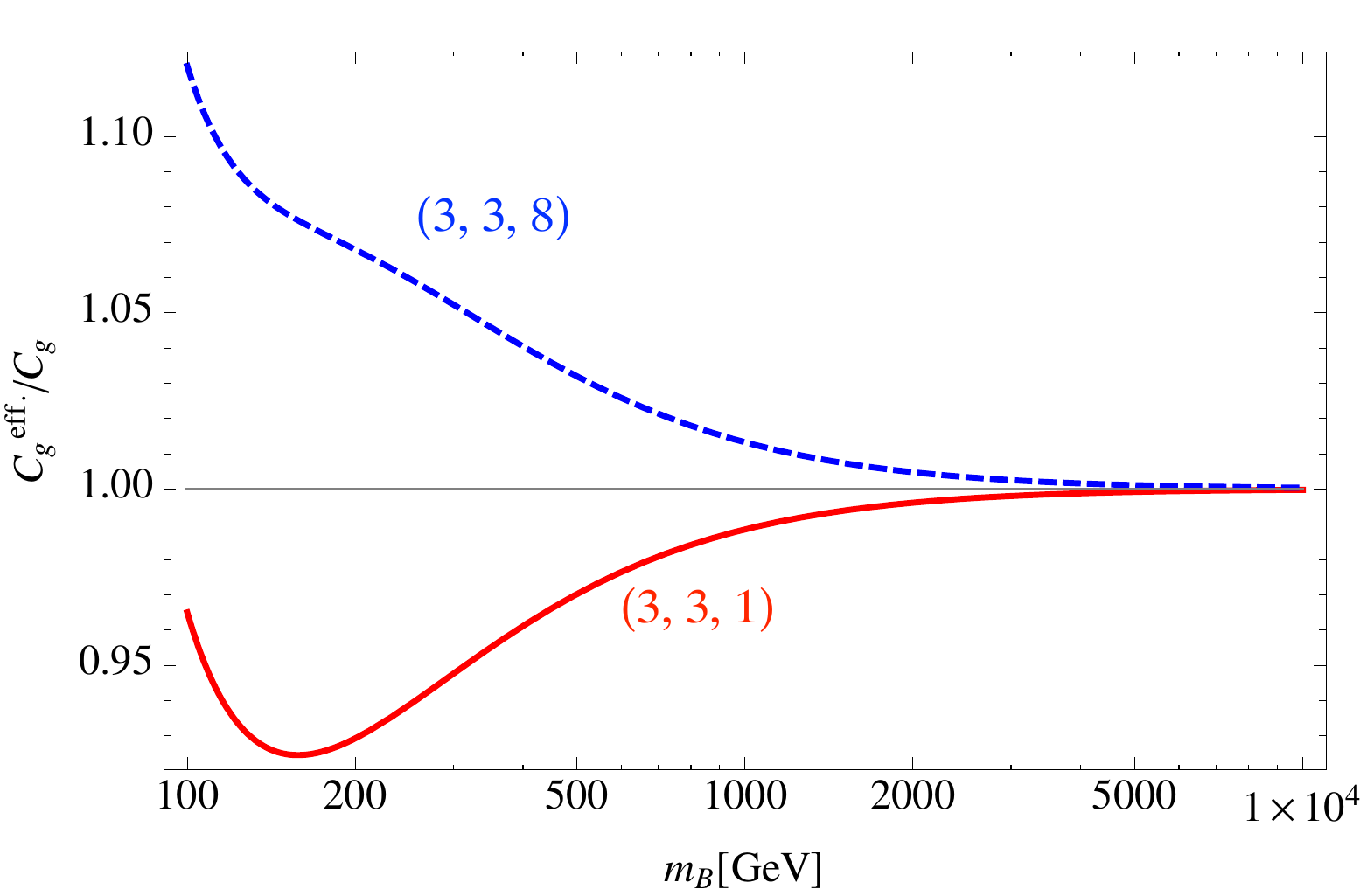}
\caption{
The ratio of $C_g^{eff.}$ and $C_g$.
The left panel is for $(A,B,S)\sim (3,1,\bar{3})$. 
In the right panel, 
the red-solid line is for $(A,B,S)\sim (3,3,1)$,
and the blue-dashed line is for $(A,B,S)\sim (3,3,8)$.
We take $m_A=100$~GeV, and $m_S = m_B$.
}
\label{fig:vsEff}
\end{figure}

We compare our full two-loop result, $C_g$,
with the result calculated based on the effective theory, $C_g^{eff.}$.
We consider three cases,
$(A,B,S) \sim (3, 1, \bar{3}),~(3, 3, 1),~(3, 3, 8)$.
We calculate the ratio of $C_g^{eff.}$ and $C_g$
by using Eqs.~\eqref{eq:our_result}, \eqref{eq:ourCGeff}, and Table.~\ref{tab:X}. 
Numerical results are shown in Fig.~\ref{fig:vsEff}.
Here we take $m_A = 100$~GeV, and $m_S = m_B$.
We find that our calculation agrees with the result in Refs.~\cite{Chang:1991ry, Chang:1992vs} very well for $m_A \ll m_B, m_S$.
This result means the validity of the calculation based on the effective theory for the large mass hierarchy regime.
On the other hand, we also find discrepancies between $C_g^{eff.}$ and $C_g$ for small mass hierarchy regime.
In particular, the discrepancy is not negligible for $(A,B,S) \sim (3, 1, \bar{3})$.
Therefore, it is better to use $C_g$ instead of $C_g^{eff.}$ for the purpose of reducing theoretical uncertainty.

Even for the large mass hierarchy regime, $C_g$ has an advantage over $C_g^{eff.}$.
Eq.~\eqref{eq:CGeff} is applicable 
only if
a lighter particle is a color non-singlet fermion.
On the other hand, Eq.~\eqref{eq:our_result} can be applied for any color representation with arbitrary mass spectra.

\section{Numerical analysis}
\label{sec:Numeric}
Let us next estimate the nucleon EDMs induced by the Weinberg operator and discuss the importance of measuring EDMs. 
The value of EDMs for proton and neutron induced by the Weinberg operator ($d_p(w)$ and $d_n(w)$) can be estimated by applying the QCD sum rules \cite{Demir:2002gg}:
\begin{align}
{d_N(w)}
=
\pm e~\Lambda_{\rm{nEDM}}~w(1~\mbox{GeV}),
~~~~(N=n,p),
\label{eq:dNmatchingw}
\end{align}
where $\Lambda_{\rm{nEDM}}=10-30$~MeV and $w(1~\mbox{GeV})$ denotes $w=g^3_sC_G$ at the scale of $1~$GeV. 
It should be noted that the factor $\Lambda_{\rm{nEDM}}$ can also be estimated by naive dimensional analysis (NDA) \cite{Manohar:1983md} and the NDA predicts a factor of two larger value than the estimation using QCD sum rules \cite{PHRVA.D41.999,PHRVA.D42.867,Chang:1992bg,Chang:2000ck,PRLTA.63.2333}. Therefore, the following phenomenological analysis may be under-estimating the hadronic uncertainty.
We also note that we cannot determine the sign of the contribution from the Weinberg operator solely from the analysis based on QCD sum rules or the NDA.

In order to estimate the values of $w(1~\mbox{GeV})$, we need to take into account the effect of the renormalization group equation (RGE) evolution between the matching energy scale ($\mu_{\rm{match}}$) and the scale of 1~GeV. The RGE for $w$ at the leading order is given as \cite{Degrassi:2005zd} 
\begin{align}
\frac{d}{d \ln\mu}w(\mu)
=
\frac{g^2_s(\mu)}{16\pi^2}(N_C+2N_f)w(\mu),
\end{align}
where $N_C=3$ and $N_f$ are the number of colors and quark flavors, respectively. Solving the RGE and plugging the solution into Eq.~(\ref{eq:dNmatchingw}), we estimate the nucleon EDMs induced by the Weinberg operator.

The Weinberg operator induces quark EDMs and cEDMs through the operator mixing effects and they also affect the nucleon EDMs. However, the contributions are subdominant compared to that from the Weinberg operator because the operator mixing effects are suppressed by the one-loop factor. For the models discussed below, we have checked $|d^{\rm{(c)EDM}}_N/d_N(w)|\lesssim \mathcal{O}(0.01)$ numerically where $d^{\rm{(c)EDM}}_N$ is the nucleon EDMs induced by quark (c)EDMs and $d_N(w)$ is the nucleon EDMs induced by Weinberg operator. Therefore, we neglect the operator mixing effects in the following analysis.

We are now ready to estimate the value of $d_N(w)$. Let us consider three cases, $(A,B,S)\sim(3,1,\overline{3}),~(3,3,1)$, and $(3,3,8)$ as concrete examples.
In the numerical evaluation of the nucleon EDMs, we use 
$
m_Z=91.1876~\mbox{GeV},~
m_t=173.1~\mbox{GeV},~
m_b=4.18~\mbox{GeV},~
m_c=1.28~\mbox{GeV},~
$
and
$
\alpha_s(m_Z)=0.1182
$
as input parameters \cite{Patrignani:2016xqp} and set the matching energy scale, $\mu_{\rm{match}}$, as the minimum value among $m_A$, $m_B$, and $m_S$ ($\mu_{\rm{match}}=\mbox{min}[m_A,m_B,m_S]$).

\begin{figure}[tb]
 \begin{center}
  \includegraphics[width=0.49\hsize]{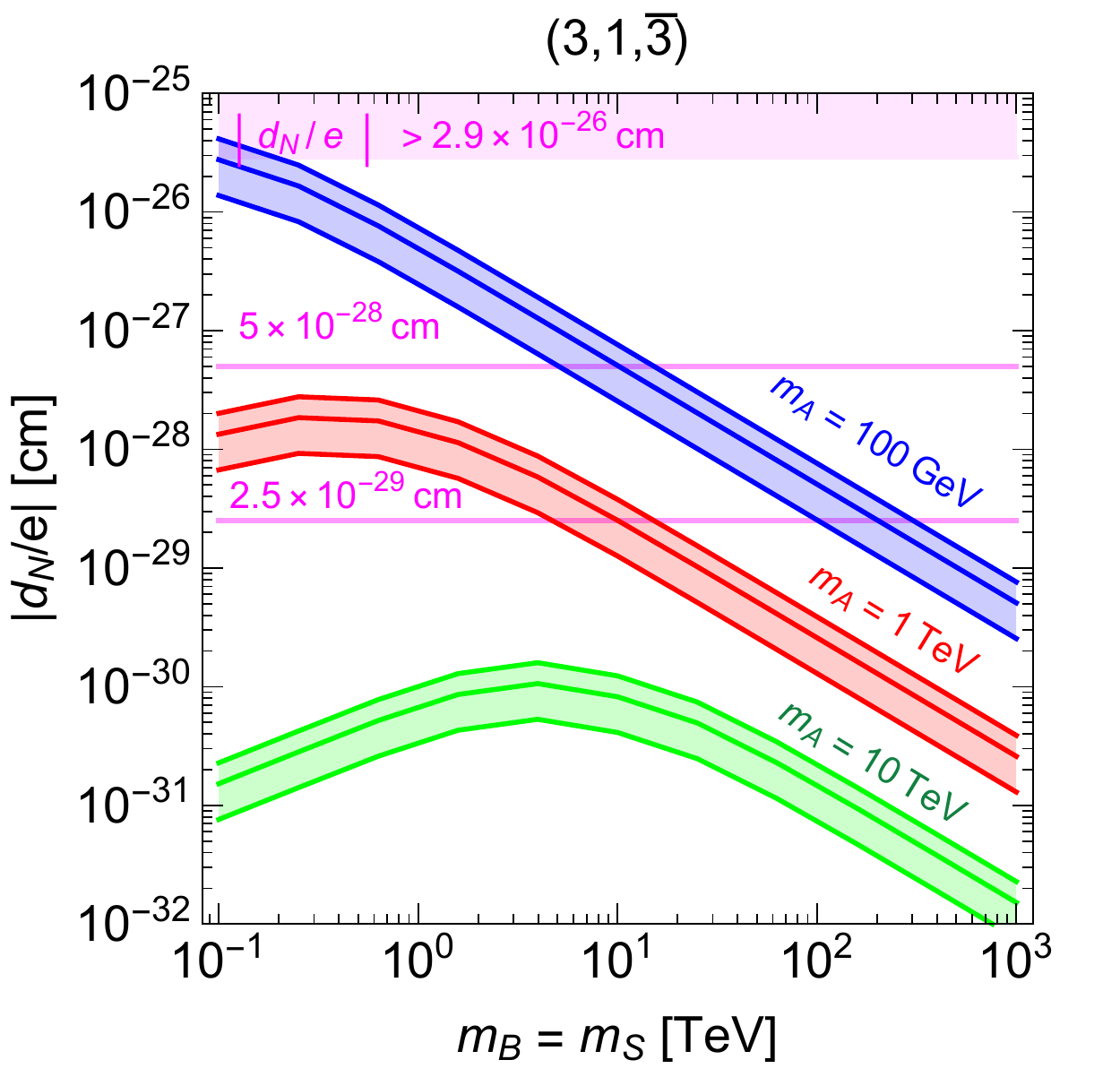}~~
  \includegraphics[width=0.49\hsize]{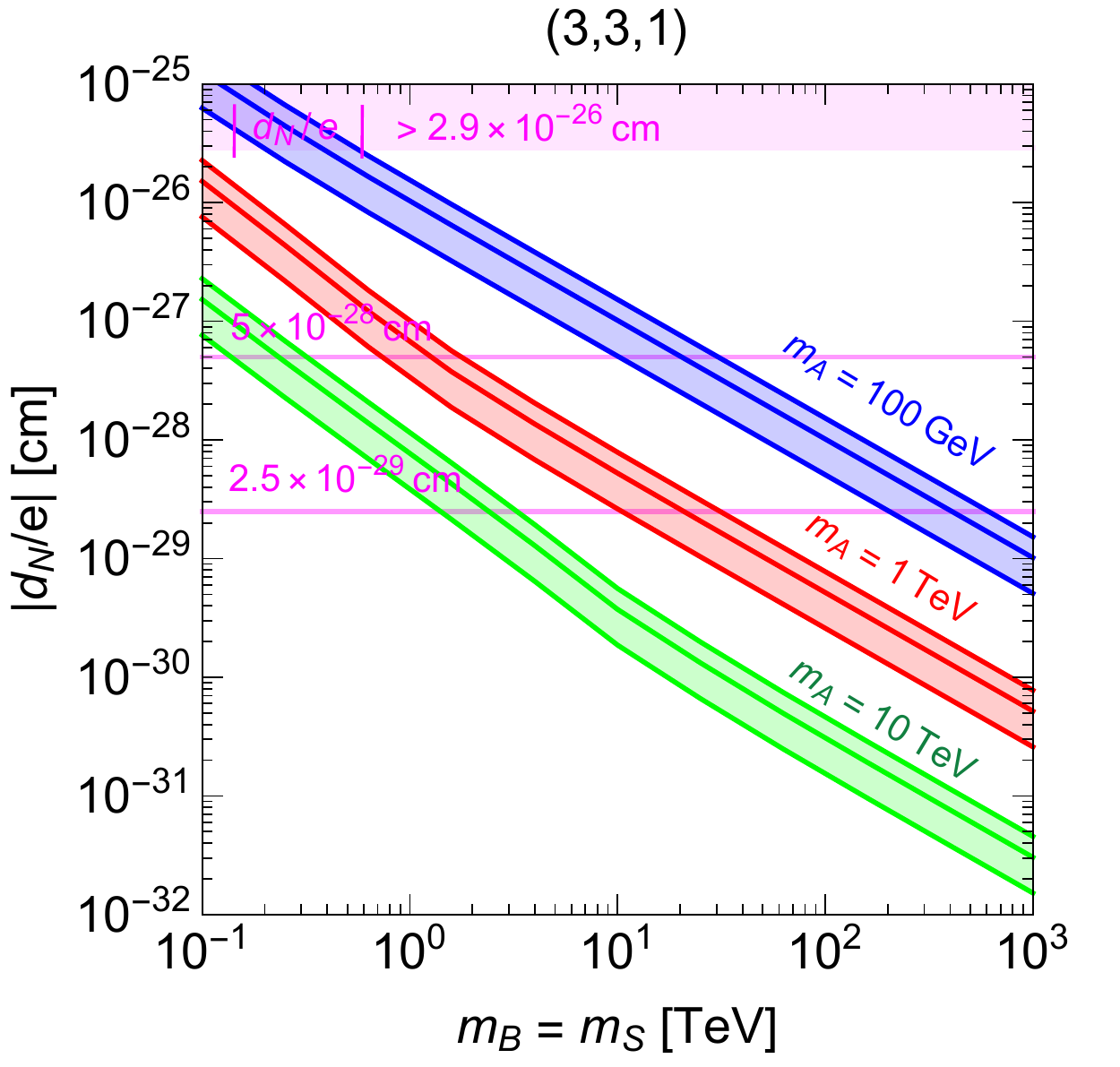}\\
  \includegraphics[width=0.49\hsize]{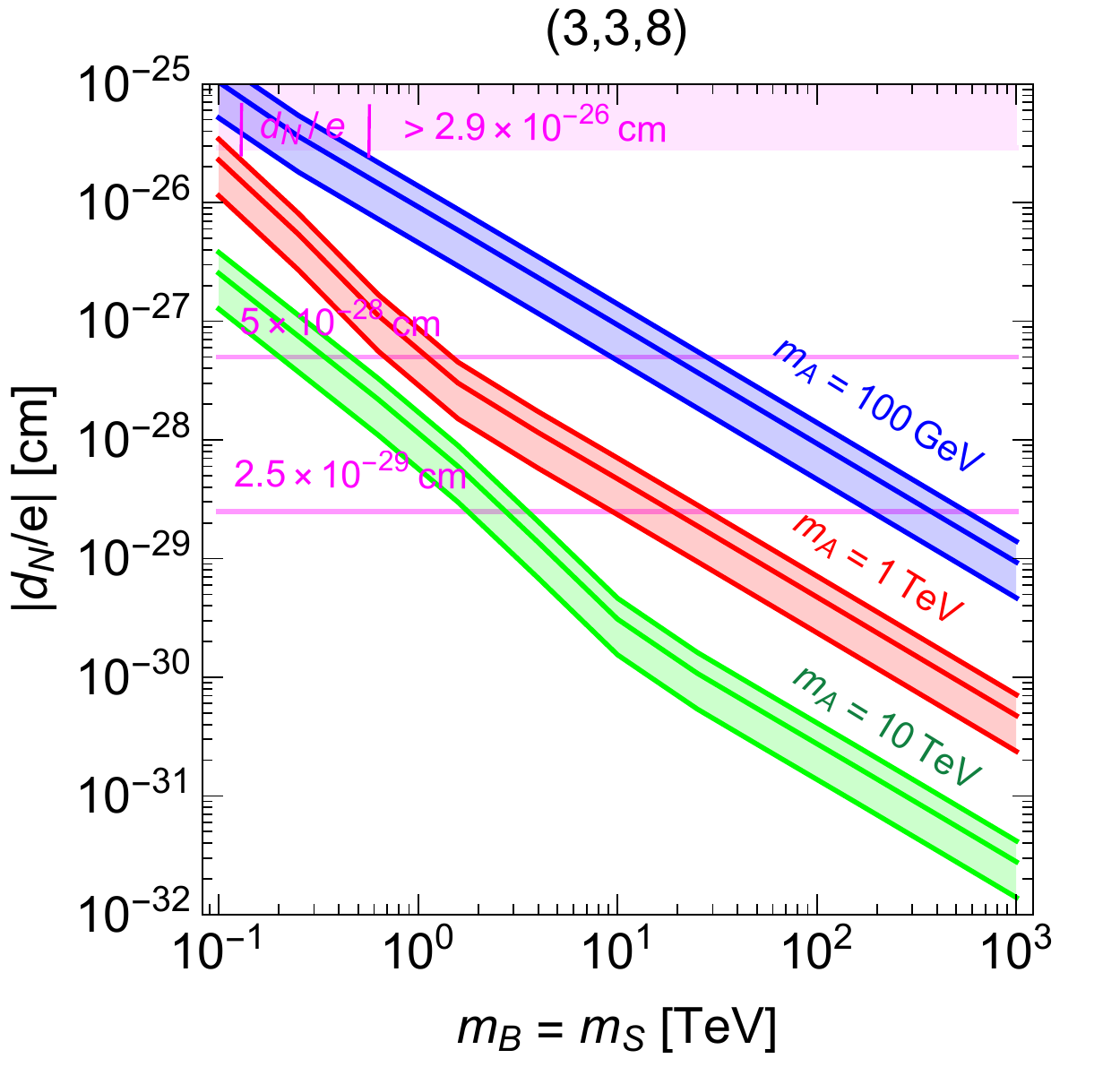}
 \end{center}
 \caption{$|d_N(w)/e|~\mbox{[cm]}$ as a function of $m_B$ [TeV] with fixing other parameters. Here we take $m_S=m_B$ and $\mbox{Im}(sa^\ast)=0.25$. Blue, red, and green lines correspond the case where $m_{A}=100$~GeV, $m_{A}=1$~TeV, and $m_{A}=10$~TeV, respectively. Magenta region corresponds the parameter region where $|d_N(w)/e|$ is larger than the current upper bound for neutron EDM \cite{hep-ex/0602020}. Magenta lines show the future prospects for the measurements of neutron and proton EDMs, $|d_n|=5\times10^{-28}~\mbox{e cm}$ \cite{Altarev:2009zz} and $|d_p|=2.5\times10^{-29}~\mbox{e cm}$ \cite{Lehrach:2012eg}.}
 \label{fig:nEDMvsm}
\end{figure}
\begin{figure}[tb]
 \begin{center}
 \includegraphics[width=0.3619\hsize]{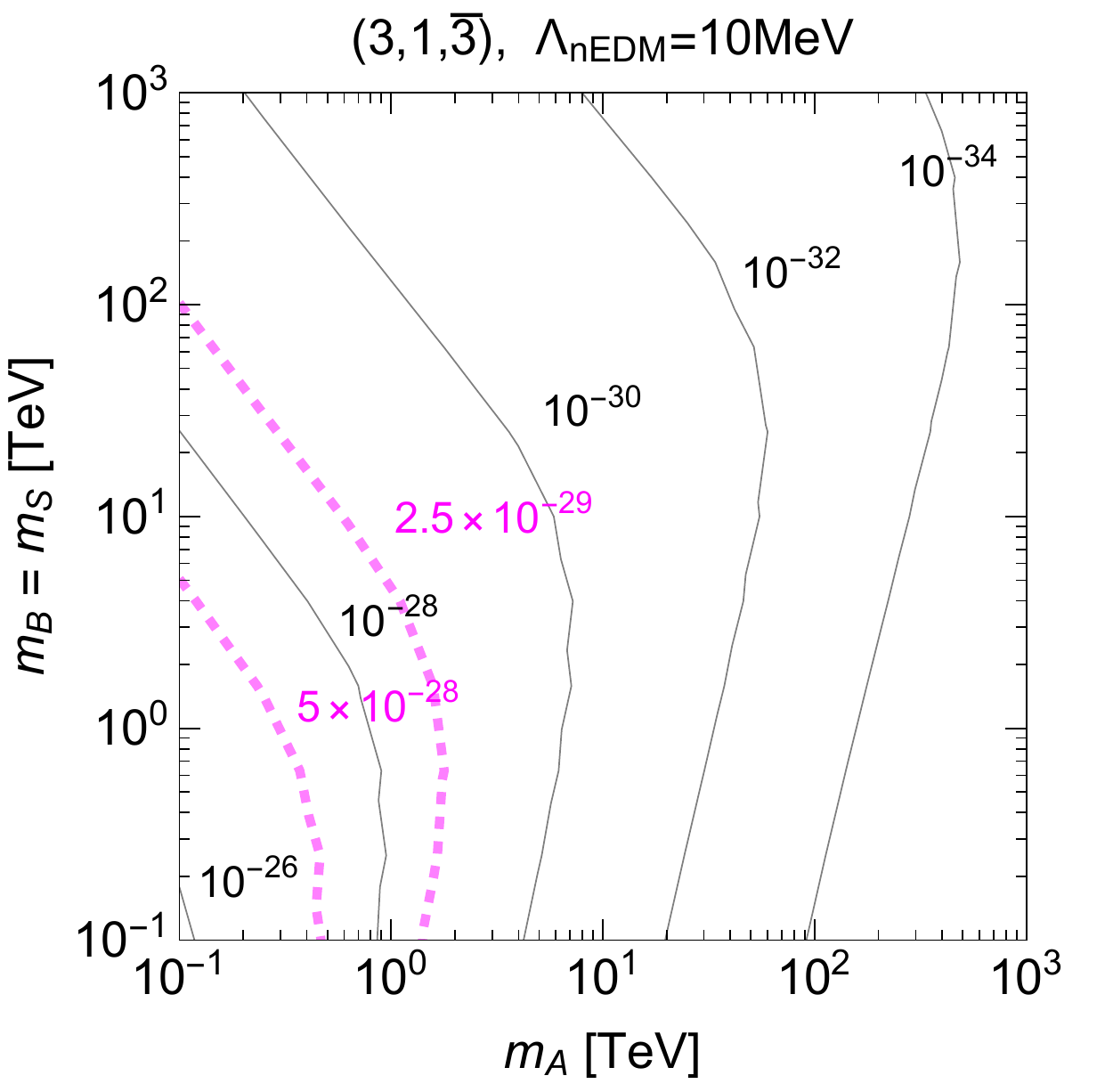}~~~
 \includegraphics[width=0.3619\hsize]{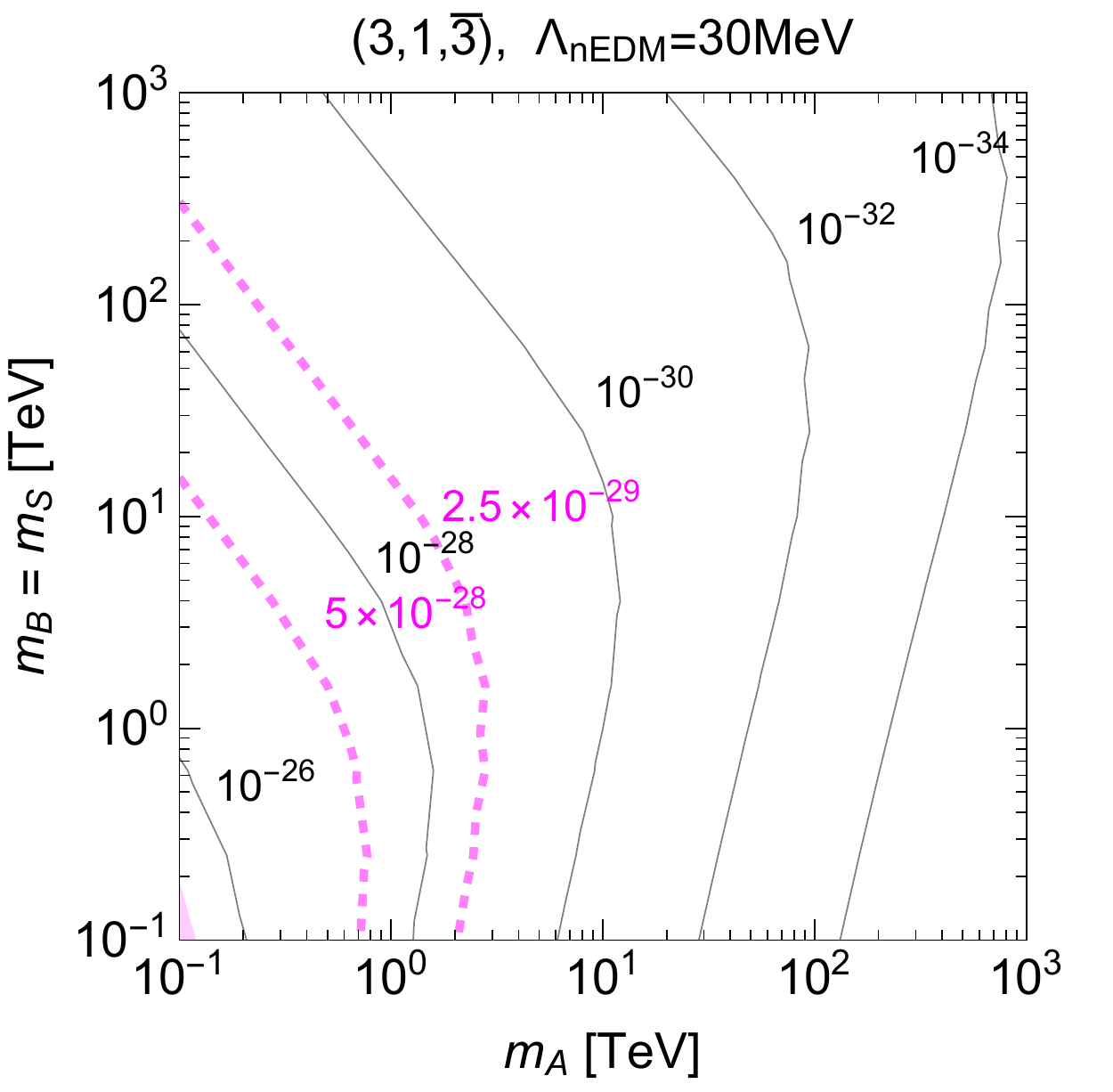}\\
 \includegraphics[width=0.3619\hsize]{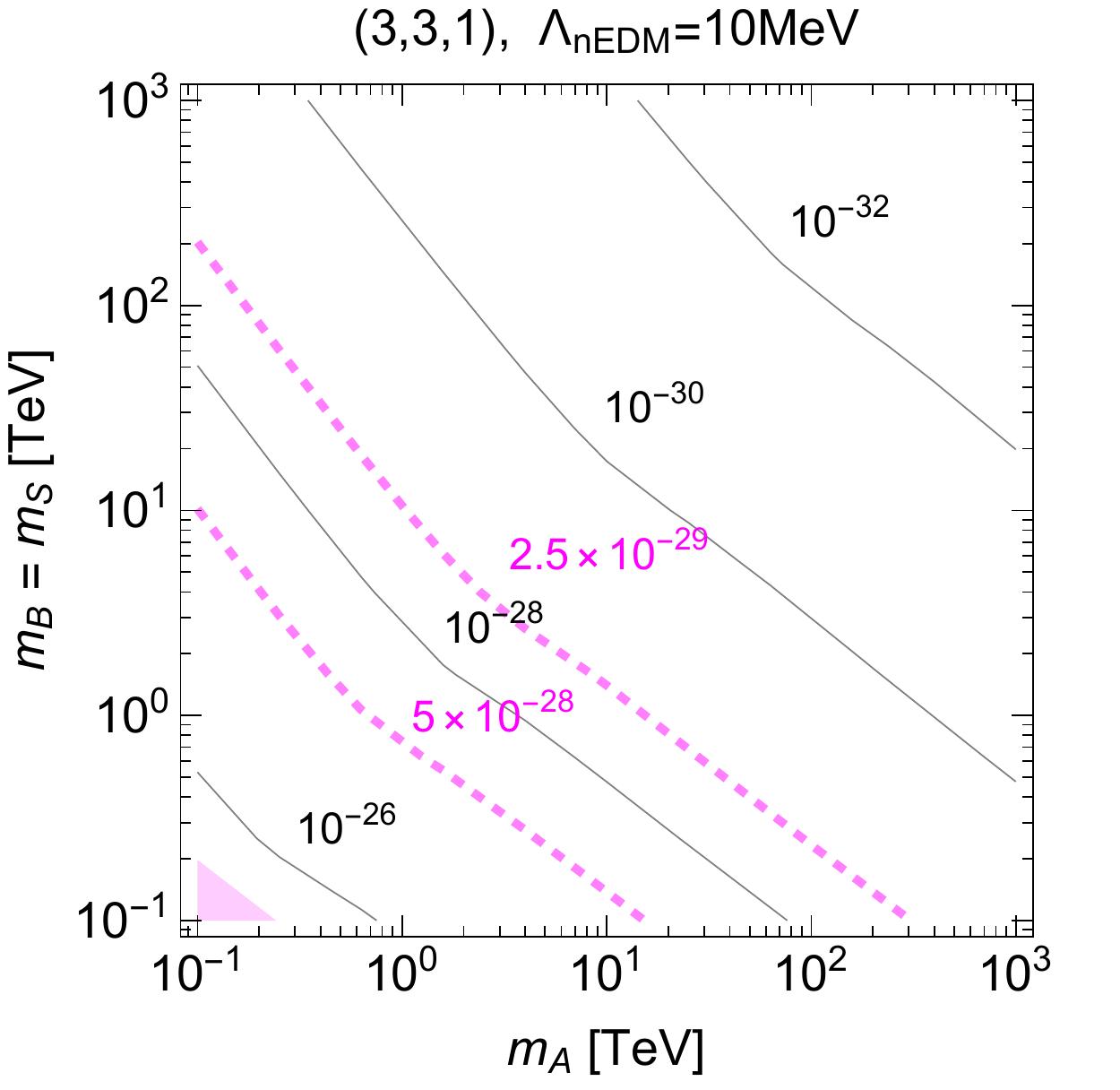}~~~
 \includegraphics[width=0.3619\hsize]{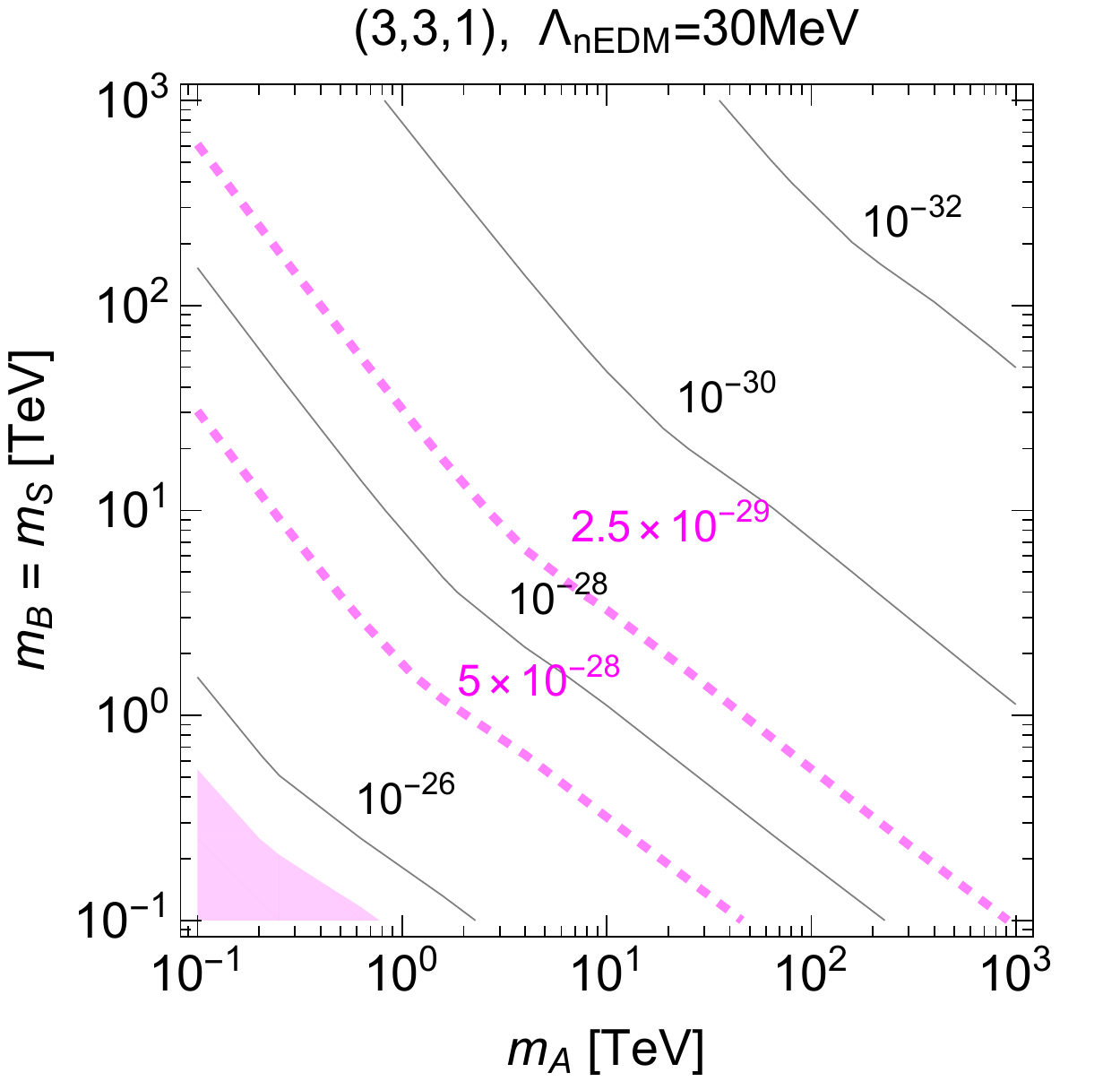}\\
 \includegraphics[width=0.3619\hsize]{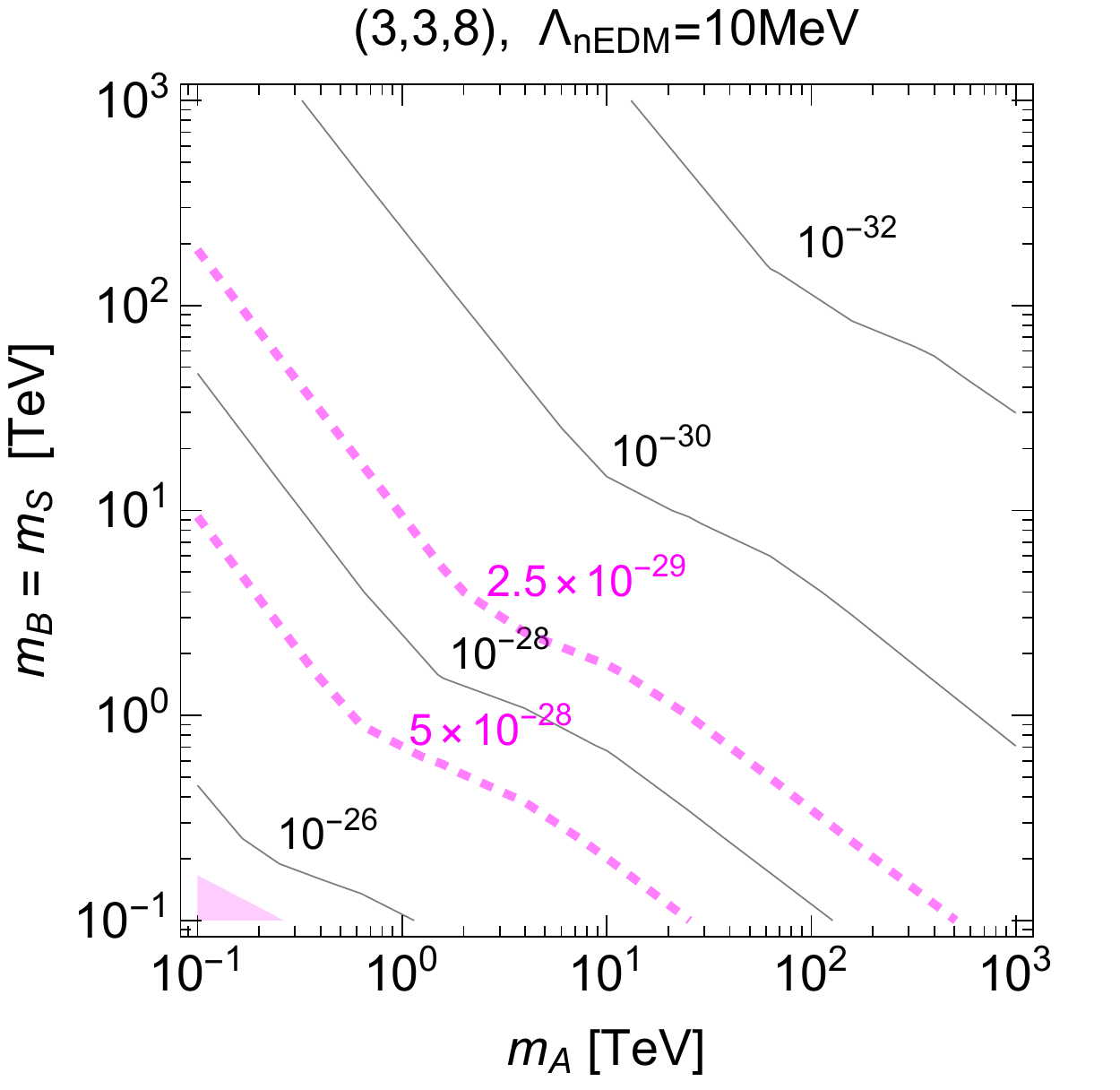}~~~
 \includegraphics[width=0.3619\hsize]{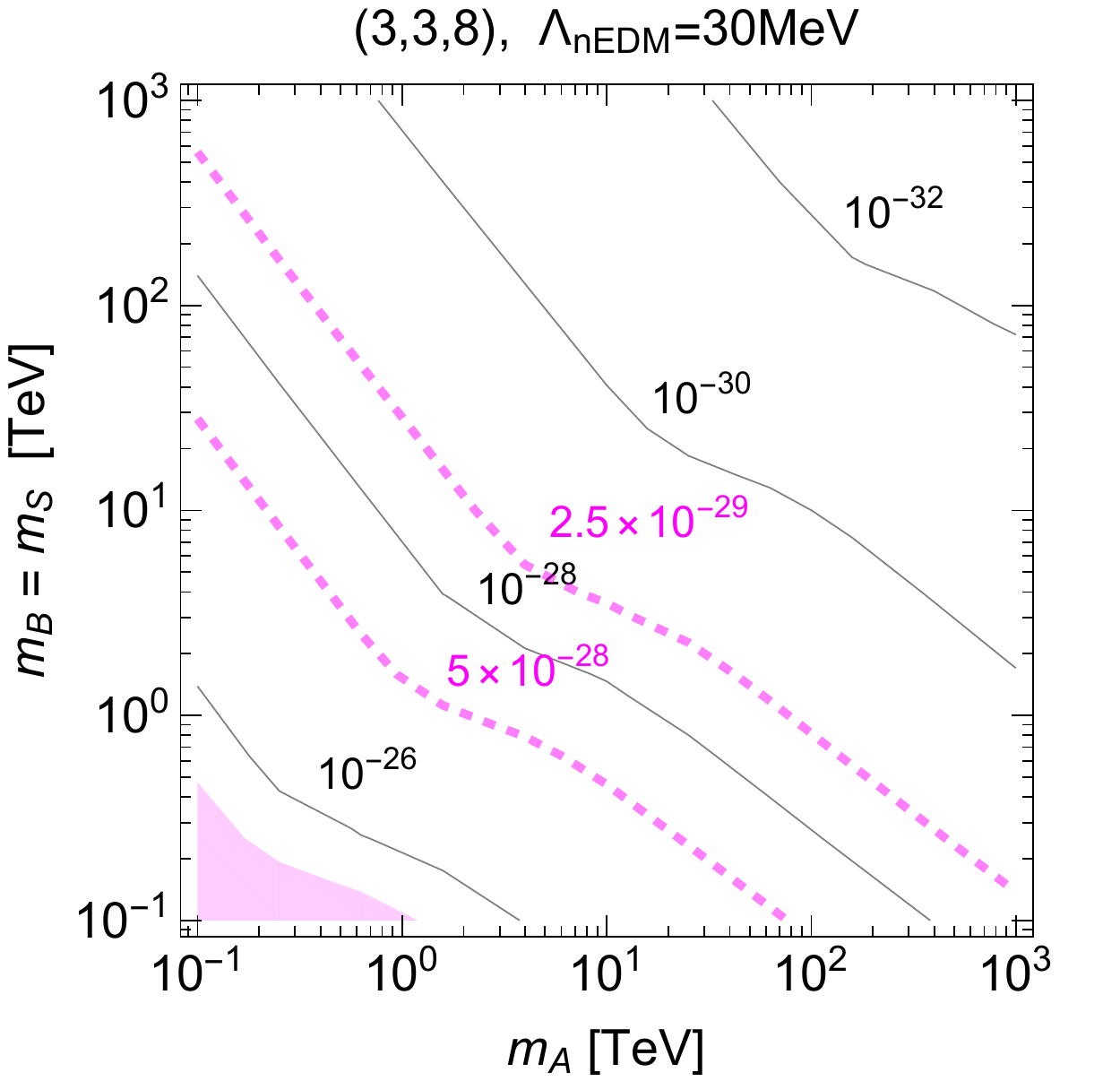}
   \end{center}
 \caption{$|d_N(w)/e|~\mbox{[cm]}$ in the ($m_A$, $m_B$) plane with fixing other parameters. Here we take $m_S=m_B$ and $\mbox{Im}(sa^\ast)=0.25$. 
In the magenta region, $|d_N(w)/e|$ is larger than the current upper bound~\cite{hep-ex/0602020},
and the magenta dashed lines show the future prospects~\cite{Altarev:2009zz, Lehrach:2012eg}.
}
 \label{fig:nEDM}
\end{figure}

Figure \ref{fig:nEDMvsm} and figure \ref{fig:nEDM} show the absolute value of $d_N(w)$ $(|d_N(w)|)$ with taking $m_S=m_B$ and $\mbox{Im}(sa^\ast)=0.25$. In figure \ref{fig:nEDMvsm}, blue, red, and green bands correspond to $|d_N(w)/e|$ in the case where $m_{A}=100$~GeV, $m_{A}=1$~TeV, and $m_{A}=10$~TeV, respectively. In these colored bands, the upper lines, middle lines, and lower lines correspond to $|d_N(w)/e|$ estimated with $\Lambda_{\rm{nEDM}}=30$~MeV, $\Lambda_{\rm{nEDM}}=20$~MeV, and $\Lambda_{\rm{nEDM}}=10$~MeV, respectively. In figure \ref{fig:nEDM}, we show the contours of $|d_N(w)/e|$ in $(m_A, m_B)$ plane. In the left panels, we take $\Lambda_{\rm{nEDM}}=10$~MeV. In the right panels, we take $\Lambda_{\rm{nEDM}}=30$~MeV.
%
%
%

In $m_A\gg m_B=m_S$ regime, we find that the mass dependence of $|d_N(w)|$ in $(A,B,S)\sim(3,1,\overline{3})$ model is different from that in the other models. In $(A,B,S)\sim(3,1,\overline{3})$ model, $|d_N(w)|$ behaves as $|d_N(w)|\propto m_B/m^3_A$ in $m_A\gg m_B=m_S$ regime. On the other hand, in the $(A,B,S)\sim(3,3,1)$ and $(3,3,8)$ models, $|d_N(w)|$ behaves as $|d_N(w)| \propto1/(m_Am_B)$ in $m_A\gg m_B=m_S$ regime. The difference of the mass dependence can be understood by taking into account the mass dependence of loop functions $f_1$ and $f_2$, which are summarized in Appendix \ref{app:fapprox}.

The region filled with magenta color in the figures shows the excluded region by the latest upper bound on neutron EDM, 
$|d_n|\leq 2.9\times 10^{-26}$~e cm \cite{hep-ex/0602020}. We find that the models with $m_A\simeq m_B\simeq m_S\simeq \mathcal{O}(100)$~GeV and $\mbox{Im}(sa^\ast)=0.25$ can be severely constrained by the current upper limit on the neutron EDM. 
For example, 
if we take $\mbox{Im}(sa^\ast)=0.25$ and $m_B=m_S=100$~GeV in $(3,3,1)$ model, 
$m_A$ should be larger than
$245-787$~GeV
to avoid the latest upper bound on neutron EDM.
The uncertainty of the lower mass bound comes from the uncertainty of the scale of $\Lambda_{\rm{nEDM}}$ in Eq.~(\ref{eq:dNmatchingw}). 

It should be noted that the existence of $\mathcal{O}(100)$~GeV colored particles might be also severely constrained by the hadron collider experiments. However, in order to estimate the collider bounds on our models explicitly, we need to specify the production and decay properties of the colored particles. The properties of colored particles highly depend on the detail of the models which is not specified in this paper. Thus, 
we defer the estimation as future work.

Magenta dashed lines show the future prospects for the observation of nucleon EDMs. Here we take $|d_n|=5\times10^{-28}~\mbox{e cm}$ \cite{Altarev:2009zz} and $|d_p|=2.5\times10^{-29}~\mbox{e cm}$ \cite{Lehrach:2012eg} as reference values of the future sensitivities. 
It is found that the future measurements for the EDMs may have sensitivities to the mass scale of $\mathcal{O}(1-10)$~TeV in each scenario. 
For example, 
if we take $\mbox{Im}(sa^\ast)=0.25$ and $m_B=m_S=1$~TeV in $(3,3,1)$ model, the future measurement of proton EDM proposed by Ref.~\cite{Lehrach:2012eg} can cover $m_A\lesssim15.6-47.1$~TeV mass region.
It is implied that future measurements of EDMs enable us to investigate physics at higher energy scale beyond the reach of collider experiments.

Before closing this section, we would like to give a comment on the uncertainty of our numerical calculation.
The main uncertainty of our numerical analysis comes from 
the QCD sum rules which we use in order to estimate the nucleon EDMs induced by the Weinberg operator. If the value of $\Lambda_{\rm{nEDM}}$ is determined more precisely, we can reduce the uncertainty of our numerical calculation. It means that the improvement of the determination $\Lambda_{\rm{nEDM}}$ gives us more detailed information of physics beyond the SM.

\section{Summary}
\label{sec:summary}

We have investigated the contribution to EDMs from a CP-violating purely gluonic dimension-6 operator
called the Weinberg operator, $GG\tilde{G}$. 
We have calculated its Wilson coefficient at the leading order in a setup 
that contains CP-violating interactions constructed by one scalar and two fermion fields.
The SU(3)$_c$ gauge invariance is imposed in the interaction terms.
The SU(3)$_c$ representations of the scalar and fermion fields are kept arbitrary in our analysis,
and thus our result can be applied to a variety of models. 
Our main result is given in Eq.~\eqref{eq:our_result}.

We have compared our full two-loop result with the result calculated based on the effective theory approach~\cite{Chang:1991ry, Chang:1992vs}.
The effective theory is applicable if the lightest particle in the loop is a colored fermion.
We have numerically shown that our result agrees with the effective theory in the region of parameter space where
the lightest particle is much lighter than the other particles.
However, if the mass difference is small, the discrepancy between two calculations are sizable 
as can be seen from Fig.~\ref{fig:vsEff}. In such a regime, we have to use the result of the full two-loop calculation.
It is worth noting that the full two-loop result given in Eq.~\eqref{eq:our_result}
is as simple as the result derived from the effective theory Eq.~\eqref{eq:ourCGeff}.

Finally, 
we have estimated the nucleon EDMs generated by the Weinberg operator, 
and have compared it with the current upper bound and future sensitivities on the neutron EDM.
We have found that all the newly introduced particles have to be heavier than ${\cal O}(100)$~GeV
due to the current upper bound on the EDM. 
We also have found that future experiments can search new particles whose masses are heavier than ${\cal O}(1)$~TeV.
For example, in $(A,B,S) \sim (3,3,1)$ model with $m_B = m_S =$1~TeV, it is possible to observe EDM 
for $m_A < $ 16--47~TeV. 
Thus the future measurements of EDMs are powerful tools for probing physics at higher energy scale beyond the reach of collider experiments.
On the other hand, the uncertainty originates from the estimation of the EDM based on the naive dimensional analysis.
It is necessary to improve the evaluation of the contribution to EDMs from the Weinberg operator
for the determination of the new physics scale by the EDM measurements.\footnote{
For a current status of a lattice calculation of the neutron EDM from the Weinberg operator, see Ref.~\cite{1711.04730}.
}

\section*{Acknowledgments}
This work was supported by JSPS KAKENHI Grant Number 16K17715 [T.A.],
16H06490 [R.N.]
and by Grant-in-Aid for Scientific research from the Ministry of Education, Science, Sports, and Culture (MEXT), Japan, No. 16H06492 [J.H.].
The work of J.H. is also supported by World Premier International Research Center Initiative (WPI Initiative), MEXT, Japan. 


\appendix
\flushleft{\textbf{\Large Appendix}}
\section{One gluon field from the scalar field in Fock-Schwinger gauge}
\label{app:GluonsFromScalar}
We briefly review the Fock-Schwinger gauge and explain 
the reason why diagrams with gluon emission from the scalar field are absent in our calculation.


We treat gluon fields as background fields obeying the Fock-Schwinger gauge,
\begin{align}
(x-x_0)^{\mu} A_{\mu}(x).  
\end{align}
This gauge fixing condition allows us to expand the gauge field by its field strength.
In practice, we need to know the Feynman rules with this background field.
Vertices with fermion propagators is modified as
\begin{align}
\frac{i}{\slashed{p} + \slashed{k} - m} 
(-ig_s \gamma^{\mu} T^a \epsilon^a_\mu(k))
\frac{i}{\slashed{p} - m} 
\to&
\frac{i}{\slashed{p} + \slashed{k} - m} 
(-ig_s \gamma^{\mu} T^a \tilde{A}^a_\mu(k))
\frac{i}{\slashed{p} - m} 
,
\end{align}
where
\begin{align}
 \tilde{A}_{\mu}(k) 
=&
\Biggl(
\frac{1}{2} G_{\alpha \mu}(x_0) \frac{1}{i} \frac{\partial}{\partial k_\alpha}
\nonumber\\
&
+
\frac{1}{3} D_{\alpha_1} G_{\alpha_2 \mu}(x_0) 
\frac{1}{i} \frac{\partial}{\partial k_{\alpha_1}}
\frac{1}{i} \frac{\partial}{\partial k_{\alpha_2}}
\nonumber\\
&
+
\frac{1}{4} \frac{1}{2!} D_{\alpha_1} D_{\alpha_2} G_{\alpha_3 \mu}(x_0) 
\frac{1}{i} \frac{\partial}{\partial k_{\alpha_1}}
\frac{1}{i} \frac{\partial}{\partial k_{\alpha_2}}
\frac{1}{i} \frac{\partial}{\partial k_{\alpha_3}}
\nonumber\\
&
+ \cdots
\Biggr) (2\pi)^4 \delta(k)
. 
\label{eq:A-tilde}
\end{align}
We also have to perform integration with respect to $k$.
Vertices with scalar propagators are similarly modified as follows.
\begin{align}
\frac{i}{(p+k)^2 - m^2} 
((2p + k)^{\mu} g_s T^a \epsilon^a_\mu(k))
\frac{i}{p^2 - m^2} 
\ \to\ &
\frac{i}{(p+k)^2 - m^2} 
((2p + k)^{\mu} g_s T^a \tilde{A}^a_\mu(k))
\frac{i}{p^2 - m^2} 
\label{eq:scalar_with_one-gluon}
,\\
\frac{i}{(p+k)^2 - m^2} 
(g_s^2 T^a T^b \epsilon^{a \mu}(k) \epsilon^b_\mu(k))
\frac{i}{p^2 - m^2} 
\ \to \ &
\frac{i}{(p+k)^2 - m^2} 
(g_s^2 T^a T^b \tilde{A}^{a \mu}(k) \tilde{A}^b_\mu(k))
\frac{i}{p^2 - m^2} 
.
\end{align}
Using these rules, we can show the absence of diagrams with gluon emission from the scalar field.

We start by considering diagrams that generate $DGD\tilde{G}$. 
In this case, we need two gluon fields, $\tilde{A}^a_{\mu}(k)$ and $\tilde{A}^b_{\nu}(k')$.
If two of them are emitted from scalar fields,
we only have two gamma matrices from the fermion propagator, and thus
cannot obtain $\epsilon$ tensor. 
Moreover, we can show that 
diagrams with a gluon emitted from a fermion field and the other gluon from the scalar field
also vanish. Let us consider a diagrams with a gluon field emitted from fermion field $A$. 
The diagram is proportional to
\begin{align}
&
\int_{p,q,k}
 \tr \left(
\gamma^5
\frac{\slashed{p}+\slashed{k} + m_A}{(p+k)^2 - m_A^2}
\gamma^{\mu}
\frac{\slashed{p} + m_A}{p^2 - m_A^2}
\frac{\slashed{p}+\slashed{k}+\slashed{q} + m_B}{(p+k+q)^2 - m_B^2}
\right)
\times (\cdots)
\nonumber\\
=&
-4i
\int_{p,q,k}
\frac{
\epsilon^{\alpha \mu \beta \sigma}
k_{\alpha} p_\beta q_{\sigma}
}{
((p+k)^2 - m_A^2)
(p^2 - m_A^2)
((p+k+q)^2 - m_B^2)
}
\times 
(\cdots)
,
\label{eq:why-one-gluon-vanish-integral}
\end{align}
where $(\cdots)$ is a term independent from $p$. 
The integral in Eq.~\eqref{eq:why-one-gluon-vanish-integral} is zero 
because of the symmetry in the $p$-integral and $\epsilon$ tensor.
Therefore, two gluon fields have to be emitted from fermion fields.
This is the reason for the absence of diagrams with gluon emission from the scalar field in Fig.~\ref{fig:Diagrams}.

We move to discuss diagrams that generate $GG\tilde{G}$. 
There are three gluon fields in this case.
We can show that at least two gluon fields have to be emitted from fermion fields 
because of the same reason in the previous paragraph.
The remains are diagrams with zero gluon fields from the scalar field and one gluon field from the scalar field.
The latter diagrams vanish thanks to a virtue of the Fock-Schwinger gauge.
The proof is the following.
From Eqs.~\eqref{eq:A-tilde} and \eqref{eq:scalar_with_one-gluon}, 
it is easy to find the scalar propagator with one gluon field strength,
\begin{align}
\frac{g_s}{2} T^a G^a_{\rho \mu}
\frac{i}{(p+k)^2 - m^2} 
\left(
(2p + k)^{\mu} \frac{\partial}{\partial k_\rho} \delta^{(4)}(k)
\right)
\frac{i}{p^2 - m^2} 
.
\end{align} 
Under the exchange of $\rho$ and $\mu$, $G_{\rho\mu}$ is anti-symmetric
but $ \delta^{(4)}(k) \frac{\partial}{\partial k_\rho} \left( [(p+k)^2 - m^2]^{-1} (2p + k)^{\mu} \right)$ 
is symmetric.
Thus we can simplify the expression as
\begin{align}
\frac{g_s}{2} T^a G^a_{\rho \mu}
\frac{i}{p^2 - m^2} 
\left(
2p^{\mu} \frac{\partial}{\partial k_\rho} \delta^{(4)}(k)
\right)
\frac{i}{p^2 - m^2} 
.
\end{align} 
This propagator vanishes if fermion propagators are independent from $k$.
The fermion propagators are independent from $k$ if four-momentum is not conserved at one of the vertices.
The violation of the momentum conservation at a vertex happens in the calculation with the Fock-Schwinger gauge
because the gauge fixing condition breaks translational invariance. 
As a result, the fermion propagators are independent from $k$, and thus the diagrams with one gluon
emission from the scalar boson vanish.
This is the reason for the absence of diagrams with gluon emission from the scalar field in Fig.~\ref{fig:Diagrams_3g}.

\section{Some formulae of integrals}
In section~\ref{sec:diagrams}, we have used the following relations.
\begin{align}
&-\frac{4}{3}
\int_\ell
\frac{\ell^4}{(\ell^2 - m_A^2)^3}
\frac{\partial^2 (B_{11} + 2 B_1 + B_0)_{(\ell^2, m_S^2, m_B^2)}}{\partial (m_B^2)^2}
=
\int_\ell
\frac{2 \ell^2 m_A^2}{(\ell^2 - m_A^2)^4}
\frac{\partial (B_{0} + B_1)_{(\ell^2, m_S^2, m_B^2)}}{\partial m_B^2}
+
(A \leftrightarrow B)
\label{eq:A1}
,\\
&\int_\ell
\frac{2 \ell^2}{(\ell^2 - m_A^2)^3} \frac{\partial^2 (B_0 + B_1)_{(\ell^2, m_S^2, m_B^2)}}{\partial (m_B^2)^2}
=
- 
\left(
\frac{1}{(\ell^2 - m_A^2)^3} 
+
\frac{3 m_A^2}{(\ell^2 - m_A^2)^4} 
\right)
 \frac{\partial B_0{}_{(\ell^2, m_S^2, m_B^2)}}{\partial m_B^2}
 +
 (A \leftrightarrow B)
\label{eq:A2}
,\\
&\int_\ell \left( \frac{2}{(\ell^2 - m_A^2 )^4} + \frac{4 m_A^2}{(\ell^2 - m_A^2 )^5} \right)
\tilde{B}_0{}_{(\ell^2, m_S^2, m_B^2)}
=
\int_\ell 
\frac{\ell^2}{(\ell^2 - m_A^2 )^4} 
\frac{\partial}{\partial \ell^2}
B_0(\ell^2, m_S^2, m_B^2)
.
\label{eq:id_for_f1}
\end{align}
To derive Eq.~\eqref{eq:id_for_f1}, we use integration by parts with
the following relation.
\begin{align}
\frac{2}{(\ell^2 - m_A^2 )^4} + \frac{4 m_A^2}{(\ell^2 - m_A^2 )^5}
=&
- \frac{1}{2}
\frac{\partial}{\partial \ell^\mu} \frac{\ell^\mu}{(\ell^2 - m_A^2 )^4} 
.
\end{align}

\section{Approximate formulae}
\label{app:fapprox}
In this appendix, we derive approximate formulae for loop functions $f_1$ and $f_2$ which can be applied in $m_A\ll m_B=m_S$ and $m_A\gg m_B=m_S$ regime. The formulae help us to understand the results given in section \ref{sec:Numeric} qualitatively.

To begin with, we focus on the case where $m_A\ll m_B=m_S$. In this regime, we find
\begin{align}
f_1(m_A,m_B,m_B)&\simeq -\frac{1}{18m^2_Am^2_B},\\
f_1(m_B,m_A,m_B)&\simeq -\frac{1}{18m^4_B},\\
f_2(m_A,m_B,m_B)&+f_2(m_B,m_A,m_B)\simeq \frac{1}{6m^2_Am^2_B}.
\end{align}
where we ignore $\mathcal{O}(m^2_A/m^2_B)$ corrections. 
Applying above expressions, we obtain
\begin{align}
C_G
&\simeq
\frac{1}{6(4\pi)^4}\text{Im}(s a^*)\frac{1}{m_Am_B},
~~\mbox{for}~~(A,B,S)\sim(3,1,\overline{3}),\\
C_G
&\simeq
-\frac{1}{3(4\pi)^4}\text{Im}(s a^*)\frac{1}{m_Am_B},
~~\mbox{for}~~(A,B,S)\sim(3,3,1),\\
C_G
&\simeq
\frac{11}{36(4\pi)^4}\text{Im}(s a^*)\frac{1}{m_Am_B},
~~\mbox{for}~~(A,B,S)\sim(3,3,8),
\end{align} 
in $m_A\ll m_B=m_S$ regime. On the other hand, in $m_A\gg m_B=m_S$ regime, we find
\begin{align}
f_1(m_A,m_B,m_B)&\simeq -\frac{1}{6m^4_A},\\
f_1(m_B,m_A,m_B)&\simeq -\frac{1}{6m^2_Am^2_B},\\
f_2(m_A,m_B,m_B)&+f_2(m_B,m_A,m_B)\simeq \frac{1}{3m^2_Am^2_B},
\end{align}
where we ignore $\mathcal{O}(m^2_B/m^2_A)$ corrections. Thus, we find that the Wilson coefficient of the Weinberg operator behaves as
\begin{align}
C_G
&\simeq
\frac{1}{2(4\pi)^4}\text{Im}(s a^*)\frac{m_B}{m_A^3},
~~\mbox{for}~~(A,B,S)\sim(3,1,\overline{3}),\\
C_G
&\simeq
-\frac{1}{2(4\pi)^4}\text{Im}(s a^*)\frac{1}{m_Am_B},
~~\mbox{for}~~(A,B,S)\sim(3,3,1),\\
C_G
&\simeq
\frac{5}{6(4\pi)^4}\text{Im}(s a^*)\frac{1}{m_Am_B},
~~\mbox{for}~~(A,B,S)\sim(3,3,8),
\end{align} 
in $m_A\gg m_B=m_S$ regime, respectively. Note that the mass dependence of $C_G$ in $(3,1,\overline{3})$ is different from the other models.

\section{A relation among $X^\dagger T_S X$, $X^\dagger T_B X$, and $X T_A T_A X^\dagger$}
\label{sec:TS_vs_TA_vs_TB}

The SU(3)$_c$ gauge invariance impose the following relations.
\begin{align}
 (U^{\dagger})_{B'B} (U)_{AA'} (U)_{SS'} X_{\bar{B} A S} =& X_{\bar{B}' A' S'},
\end{align}
where $U_{AA'}$, $U_{SS'}$, and $U^\dagger_{B'B}$ are the unitary matrices 
under the SU(3)$_c$ gauge transformation
for the fields $A$, $S$, and $B$, respectively.
Since $X$ consists of $\epsilon$ tensors and Kronecker deltas, $X$ is an invariant tensor,
and thus 
\begin{align}
 (T^a)_{AA'} X_{\bar{B'} A S'} +  (T^a)_{SS'} X_{\bar{B'} A' S} - (T^a)_{B'B} X_{\bar{B} A' S'}
=0.
\end{align}
Using this relation, we find
\begin{align}
 (X^{\dagger} T_S X) (T^a)_{AA'}
=&
 (X^{\dagger} T_B X) (T^a)_{AA'}
 - X^{\dagger}_{\bar{A} B \bar{S}} (T^a) _{A'' A} X_{\bar{B}A''S}
.
\end{align}
It is easy to show that 
\begin{align}
  X^{\dagger}_{\bar{A} B \bar{S}} (T^a) _{A'' A'} X_{\bar{B}A''S}
=&
 \frac{1}{N(r_A)} (X T_A T_A X^\dagger) (T^a)_{AA'},
\end{align} 
where $N(r_A)$ is defined through $\tr(T^a T^b) = N(r_A) \delta^{ab}$.
Finally, we find Eq.~\eqref{eq:TS_vs_TA_vs_TB}.


\end{document}